\documentstyle{cargese}%----------------------------------------------
\def\fileversion{v1.20a}% was \def\fileversion{v1.20}%
\def\filedate{21.6.94}%  was \def\filedate{26.1.94}%
\edef\epsfigRestoreAt{\catcode`@=\number\catcode`@\relax}%
\catcode`\@=11\relax
\ifx\undefined\@makeother                % -pks-
\def\@makeother#1{\catcode`#1=12\relax}  % -pks-
\fi                                      % -pks-
\immediate\write16{Document style option `epsfig', \fileversion\space
<\filedate> (edited by SPQR + pks)}% was <\filedate> (edited by SPQR)}%
\newcount\EPS@Height \newcount\EPS@Width \newcount\EPS@xscale
\newcount\EPS@yscale
\def\psfigdriver#1{%
  \bgroup\edef\next{\def\noexpand\tempa{#1}}%
    \uppercase\expandafter{\next}%
    \def\LN{DVITOLN03}%
    \def\DVItoPS{DVITOPS}%
    \def\DVIPS{DVIPS}%
    \def\emTeX{EMTEX}%
    \def\OzTeX{OZTEX}%
    \def\Textures{TEXTURES}%
    \global\chardef\fig@driver=0
    \ifx\tempa\LN
        \global\chardef\fig@driver=0\fi
    \ifx\tempa\DVItoPS
        \global\chardef\fig@driver=1\fi
    \ifx\tempa\DVIPS
        \global\chardef\fig@driver=2\fi
    \ifx\tempa\emTeX
        \global\chardef\fig@driver=3\fi
    \ifx\tempa\OzTeX
        \global\chardef\fig@driver=4\fi
    \ifx\tempa\Textures
        \global\chardef\fig@driver=5\fi
  \egroup
\def\psfig@start{}%
\def\psfig@end{}%
\def\epsfig@gofer{}%
\ifcase\fig@driver
\typeout{WARNING! ****
 no specials for LN03 psfig}%
\or % case 1: dvitops
\def\psfig@start{}%
\def\psfig@end{\special{dvitops: import \@p@sfilefinal \space
\@p@swidth sp \space \@p@sheight sp \space fill}%
\if@clip \typeout{Clipping not supported}\fi
\if@angle \typeout{Rotating not supported}\fi
}%
\let\epsfig@gofer\psfig@end
\or %case2 dvips
\def\psfig@start{\special{ps::[begin]  \@p@swidth \space \@p@sheight \space%
        \@p@sbbllx \space \@p@sbblly \space%
        \@p@sbburx \space \@p@sbbury \space%
        startTexFig \space }%
        \if@clip
                \if@verbose
                        \typeout{(clipped to BB) }%
                \fi
                \special{ps:: doclip \space }%
        \fi
        \if@angle              % moved after \if@clip ... \fi -pks-
                \special {ps:: \@p@sangle \space rotate \space}
        \fi
        \special{ps: plotfile \@p@sfilefinal \space }%
        \special{ps::[end] endTexFig \space }%
}%
\def\psfig@end{}%
\def\epsfig@gofer{\if@clip
                        \if@verbose
                           \typeout{(clipped to BB)}%
                        \fi
                        \epsfclipon
                  \fi
                  \epsfsetgraph{\@p@sfilefinal}%
}%
\or % case 3, emTeX
\typeout{WARNING. You must have a .bb info file with the Bounding Box
  of the pcx file}%
\def\psfig@start{}%
\def\psfig@end{\typeout{pcx import of \@p@sfilefinal}%
\if@clip \typeout{Clipping not supported}\fi
\if@angle \typeout{Rotating not supported}\fi
\raisebox{\@p@srheight sp}{\special{em: graph \@p@sfilefinal}}}%
\def\epsfig@gofer{}%
\or % case 4, OzTeX
\def\psfig@start{}%
\def\psfig@end{%
\EPS@Width\@p@swidth
\EPS@Height\@p@sheight
\divide\EPS@Width by 65781  % convert sp to bp
\divide\EPS@Height by 65781
\special{epsf=\@p@sfilefinal
\space
width=\the\EPS@Width
\space
height=\the\EPS@Height
}%
\if@clip \typeout{Clipping not supported}\fi
\if@angle \typeout{Rotating not supported}\fi
}%
\let\epsfig@gofer\psfig@end
\or % case 5, Textures
\def\psfig@end{
         \EPS@Width=\@bbw  
         \divide\EPS@Width by 1000
         \EPS@xscale=\@p@swidth \divide \EPS@xscale by \EPS@Width
         \EPS@Height=\@bbh  
         \divide\EPS@Height by 1000
         \EPS@yscale=\@p@sheight \divide \EPS@yscale by\EPS@Height
  \ifnum\EPS@xscale>\EPS@yscale\EPS@xscale=\EPS@yscale\fi
\if@clip
   \if@verbose
      \typeout{(clipped to BB)}%
   \fi
   \epsfclipon
\fi
\special{illustration \@p@sfilefinal\space scaled \the\EPS@xscale}%
}%
\def\psfig@start{}%
\let\epsfig\psfig
\else
\typeout{WARNING. *** unknown  driver - no psfig}%
\fi
}%
\newdimen\ps@dimcent
\ifx\undefined\fbox
\newdimen\fboxrule
\newdimen\fboxsep
\newdimen\ps@tempdima
\newbox\ps@tempboxa
\long\def\fbox#1{\leavevmode\setbox\ps@tempboxa\hbox{#1}\ps@tempdima\fboxrule
    \advance\ps@tempdima \fboxsep \advance\ps@tempdima \dp\ps@tempboxa
   \hbox{\lower \ps@tempdima\hbox
  {\vbox{\hrule height \fboxrule
          \hbox{\vrule width \fboxrule \hskip\fboxsep
          \vbox{\vskip\fboxsep \box\ps@tempboxa\vskip\fboxsep}\hskip
                 \fboxsep\vrule width \fboxrule}%
                 \hrule height \fboxrule}}}}%
\fi
\ifx\@ifundefined\undefined
\long\def\@ifundefined#1#2#3{\expandafter\ifx\csname
  #1\endcsname\relax#2\else#3\fi}%
\fi
\@ifundefined{typeout}%
{\gdef\typeout#1{\immediate\write\sixt@@n{#1}}}%
{\relax}%
\@ifundefined{epsfig}{}{\typeout{EPSFIG --- already loaded} }%
\@ifundefined{epsfbox}{\input epsf}{}%
\ifx\undefined\@latexerr
        \newlinechar`\^^J
        \def\@spaces{\space\space\space\space}%
        \def\@latexerr#1#2{%
        \edef\@tempc{#2}\expandafter\errhelp\expandafter{\@tempc}%
        \typeout{Error. \space see a manual for explanation.^^J
         \space\@spaces\@spaces\@spaces Type \space H <return> \space for
         immediate help.}\errmessage{#1}}%
\fi
\def\@whattodo{You tried to include a PostScript figure which
cannot be found^^JIf you press return to carry on anyway,^^J
The failed name will be printed in place of the figure.^^J
or type X to quit}%
\def\@whattodobb{You tried to include a PostScript figure which
has no^^Jbounding box, and you supplied none.^^J
If you press return to carry on anyway,^^J
The failed name will be printed in place of the figure.^^J
or type X to quit}%
\def\@nnil{\@nil}%
\def\@empty{}%
\def\@psdonoop#1\@@#2#3{}%
\def\@psdo#1:=#2\do#3{\edef\@psdotmp{#2}\ifx\@psdotmp\@empty \else
    \expandafter\@psdoloop#2,\@nil,\@nil\@@#1{#3}\fi}%
\def\@psdoloop#1,#2,#3\@@#4#5{\def#4{#1}\ifx #4\@nnil \else
       #5\def#4{#2}\ifx #4\@nnil \else#5\@ipsdoloop #3\@@#4{#5}\fi\fi}%
\def\@ipsdoloop#1,#2\@@#3#4{\def#3{#1}\ifx #3\@nnil
       \let\@nextwhile=\@psdonoop \else
      #4\relax\let\@nextwhile=\@ipsdoloop\fi\@nextwhile#2\@@#3{#4}}%
\def\@tpsdo#1:=#2\do#3{\xdef\@psdotmp{#2}\ifx\@psdotmp\@empty \else
    \@tpsdoloop#2\@nil\@nil\@@#1{#3}\fi}%
\def\@tpsdoloop#1#2\@@#3#4{\def#3{#1}\ifx #3\@nnil
       \let\@nextwhile=\@psdonoop \else
      #4\relax\let\@nextwhile=\@tpsdoloop\fi\@nextwhile#2\@@#3{#4}}%
\long\def\epsfaux#1#2:#3\\{\ifx#1\epsfpercent
   \def\testit{#2}\ifx\testit\epsfbblit
        \@atendfalse
        \epsf@atend #3 . \\%
        \if@atend
           \if@verbose
                \typeout{epsfig: found `(atend)'; continuing search}%
           \fi
        \else
                \epsfgrab #3 . . . \\%
                \epsffileokfalse\global\no@bbfalse
                \global\epsfbbfoundtrue
        \fi
   \fi\fi}%
\def\epsf@atendlit{(atend)}
\def\epsf@atend #1 #2 #3\\{%
   \def\epsf@tmp{#1}\ifx\epsf@tmp\empty
      \epsf@atend #2 #3 .\\\else
   \ifx\epsf@tmp\epsf@atendlit\@atendtrue\fi\fi}%

\chardef\trig@letter = 11
\chardef\other = 12
 
\newif\ifdebug %%% turn me on to see TeX hard at work ...
\newif\ifc@mpute %%% don't need to compute some values
\newif\if@atend
\c@mputetrue % but assume that we do
 
\let\then = \relax
\def\r@dian{pt }%
\let\r@dians = \r@dian
\let\dimensionless@nit = \r@dian
\let\dimensionless@nits = \dimensionless@nit
\def\internal@nit{sp }%
\let\internal@nits = \internal@nit
\newif\ifstillc@nverging
\def \Mess@ge #1{\ifdebug \then \message {#1} \fi}%
 
{ %%% Things that need abnormal catcodes %%%
        \catcode `\@ = \trig@letter
        \gdef \nodimen {\expandafter \n@dimen \the \dimen}%
        \gdef \term #1 #2 #3%
               {\edef \t@ {\the #1}%%% freeze parameter 1 (count, by value)
                \edef \t@@ {\expandafter \n@dimen \the #2\r@dian}%
                \t@rm {\t@} {\t@@} {#3}%
               }%
        \gdef \t@rm #1 #2 #3%
               {{%
                \count 0 = 0
                \dimen 0 = 1 \dimensionless@nit
                \dimen 2 = #2\relax
                \Mess@ge {Calculating term #1 of \nodimen 2}%
                \loop
                \ifnum  \count 0 < #1
                \then   \advance \count 0 by 1
                        \Mess@ge {Iteration \the \count 0 \space}%
                        \Multiply \dimen 0 by {\dimen 2}%
                        \Mess@ge {After multiplication, term = \nodimen 0}%
                        \Divide \dimen 0 by {\count 0}%
                        \Mess@ge {After division, term = \nodimen 0}%
                \repeat
                \Mess@ge {Final value for term #1 of
                                \nodimen 2 \space is \nodimen 0}%
                \xdef \Term {#3 = \nodimen 0 \r@dians}%
                \aftergroup \Term
               }}%
        \catcode `\p = \other
        \catcode `\t = \other
        \gdef \n@dimen #1pt{#1} %%% throw away the ``pt''
}%
 
\def \Divide #1by #2{\divide #1 by #2} %%% just a synonym
 
\def \Multiply #1by #2%%% allows division of a dimen by a dimen
       {{%%% should really freeze parameter 2 (dimen, passed by value)
        \count 0 = #1\relax
        \count 2 = #2\relax
        \count 4 = 65536
        \Mess@ge {Before scaling, count 0 = \the \count 0 \space and
                        count 2 = \the \count 2}%
        \ifnum  \count 0 > 32767 %%% do our best to avoid overflow
        \then   \divide \count 0 by 4
                \divide \count 4 by 4
        \else   \ifnum  \count 0 < -32767
                \then   \divide \count 0 by 4
                        \divide \count 4 by 4
                \else
                \fi
        \fi
        \ifnum  \count 2 > 32767 %%% while retaining reasonable accuracy
        \then   \divide \count 2 by 4
                \divide \count 4 by 4
        \else   \ifnum  \count 2 < -32767
                \then   \divide \count 2 by 4
                        \divide \count 4 by 4
                \else
                \fi
        \fi
        \multiply \count 0 by \count 2
        \divide \count 0 by \count 4
        \xdef \product {#1 = \the \count 0 \internal@nits}%
        \aftergroup \product
       }}%
 
\def\r@duce{\ifdim\dimen0 > 90\r@dian \then   % sin(x) = sin(180-x)
                \multiply\dimen0 by -1
                \advance\dimen0 by 180\r@dian
                \r@duce
            \else \ifdim\dimen0 < -90\r@dian \then  % sin(x) = sin(360+x)
                \advance\dimen0 by 360\r@dian
                \r@duce
                \fi
            \fi}%
 
\def\Sine#1%
       {{%
        \dimen 0 = #1 \r@dian
        \r@duce
        \ifdim\dimen0 = -90\r@dian \then
           \dimen4 = -1\r@dian
           \c@mputefalse
        \fi
        \ifdim\dimen0 = 90\r@dian \then
           \dimen4 = 1\r@dian
           \c@mputefalse
        \fi
        \ifdim\dimen0 = 0\r@dian \then
           \dimen4 = 0\r@dian
           \c@mputefalse
        \fi
        \ifc@mpute \then
                \divide\dimen0 by 180
                \dimen0=3.141592654\dimen0
                \dimen 2 = 3.1415926535897963\r@dian %%% a well-known constant
                \divide\dimen 2 by 2 %%% we only deal with -pi/2 : pi/2
                \Mess@ge {Sin: calculating Sin of \nodimen 0}%
                \count 0 = 1 %%% see power-series expansion for sine
                \dimen 2 = 1 \r@dian %%% ditto
                \dimen 4 = 0 \r@dian %%% ditto
                \loop
                        \ifnum  \dimen 2 = 0 %%% then we've done
                        \then   \stillc@nvergingfalse
                        \else   \stillc@nvergingtrue
                        \fi
                        \ifstillc@nverging %%% then calculate next term
                        \then   \term {\count 0} {\dimen 0} {\dimen 2}%
                                \advance \count 0 by 2
                                \count 2 = \count 0
                                \divide \count 2 by 2
                                \ifodd  \count 2 %%% signs alternate
                                \then   \advance \dimen 4 by \dimen 2
                                \else   \advance \dimen 4 by -\dimen 2
                                \fi
                \repeat
        \fi
                        \xdef \sine {\nodimen 4}%
       }}%
 
\def\Cosine#1{\ifx\sine\UnDefined\edef\Savesine{\relax}\else
                             \edef\Savesine{\sine}\fi
        {\dimen0=#1\r@dian\multiply\dimen0 by -1
         \advance\dimen0 by 90\r@dian
         \Sine{\nodimen 0}%
         \xdef\cosine{\sine}%
         \xdef\sine{\Savesine}}}
\def\psdraft{\def\@psdraft{0}}%
\def\psfull{\def\@psdraft{1}}%
\psfull
\newif\if@compress
\def\pscompress{\@compresstrue}
\def\psnocompress{\@compressfalse}
\@compressfalse
\newif\if@scalefirst
\def\psscalefirst{\@scalefirsttrue}%
\def\psrotatefirst{\@scalefirstfalse}%
\psrotatefirst
\newif\if@draftbox
\def\psnodraftbox{\@draftboxfalse}%
\@draftboxtrue
\newif\if@noisy
\@noisyfalse
\newif\ifno@bb
\newif\if@bbllx
\newif\if@bblly
\newif\if@bburx
\newif\if@bbury
\newif\if@height
\newif\if@width
\newif\if@rheight
\newif\if@rwidth
\newif\if@angle
\newif\if@clip
\newif\if@verbose
\newif\if@prologfile
\def\@p@@sprolog#1{\@prologfiletrue\def\@prologfileval{#1}}%
\def\@p@@sclip#1{\@cliptrue}%
\newif\ifepsfig@dos  % only single suffix possible
\def\epsfigdos{\epsfig@dostrue}%
\epsfig@dosfalse
\newif\ifuse@psfig
\def\ParseName#1{\expandafter\@Parse#1}%
\def\@Parse#1.#2:{\gdef\BaseName{#1}\gdef\FileType{#2}}%

\def\@p@@sfile#1{%
  \ifepsfig@dos
     \ParseName{#1:}%
  \else
     \gdef\BaseName{#1}\gdef\FileType{}%
  \fi
  \def\@p@sfile{NO FILE: #1}%
  \def\@p@sfilefinal{NO FILE: #1}%
  \openin1=#1
  \ifeof1\closein1\openin1=\BaseName.bb
    \ifeof1\closein1
      \if@bbllx                 % No postscript file but bb given explicitly.
        \if@bblly\if@bburx\if@bbury
          \def\@p@sfile{#1}%
          \def\@p@sfilefinal{#1}%
        \fi\fi\fi
      \else                     % No bounding box found.
        \@latexerr{ERROR. PostScript file #1 not found}\@whattodo
        \@p@@sbbllx{100bp}%
        \@p@@sbblly{100bp}%
        \@p@@sbburx{200bp}%
        \@p@@sbbury{200bp}%
        \psdraft
      \fi
    \else                       % Postscript file is compressed.
      \closein1%
      \edef\@p@sfile{\BaseName.bb}%
      \typeout{using BB from \@p@sfile}%
      \ifnum\fig@driver=3
        \edef\@p@sfilefinal{\BaseName.pcx}%
      \else
        \ifepsfig@dos
          \edef\@p@sfilefinal{"`gunzip -c `texfind \BaseName.{z,Z,gz}"}%
        \else
          \edef\@p@sfilefinal{"`epsfig \if@compress-c \fi#1"}%          
        \fi
      \fi
    \fi
  \else\closein1                % Postscript file is not compressed.
    \edef\@p@sfile{#1}%
    \if@compress  
      \edef\@p@sfilefinal{"`epsfig -c #1"}%
    \else
      \edef\@p@sfilefinal{#1}%
    \fi
  \fi%
}

\let\@p@@sfigure\@p@@sfile
\def\@p@@sbbllx#1{%
                                            \@bbllxtrue
                \ps@dimcent=#1
                \edef\@p@sbbllx{\number\ps@dimcent}%
                \divide\ps@dimcent by65536
                \global\edef\epsfllx{\number\ps@dimcent}%
}%
\def\@p@@sbblly#1{%
                \@bbllytrue
                \ps@dimcent=#1
                \edef\@p@sbblly{\number\ps@dimcent}%
                \divide\ps@dimcent by65536
                \global\edef\epsflly{\number\ps@dimcent}%
}%
\def\@p@@sbburx#1{%
                \@bburxtrue
                \ps@dimcent=#1
                \edef\@p@sbburx{\number\ps@dimcent}%
                \divide\ps@dimcent by65536
                \global\edef\epsfurx{\number\ps@dimcent}%
}%
\def\@p@@sbbury#1{%
                \@bburytrue
                \ps@dimcent=#1
                \edef\@p@sbbury{\number\ps@dimcent}%
                \divide\ps@dimcent by65536
                \global\edef\epsfury{\number\ps@dimcent}%
}%
\def\@p@@sheight#1{%
                \@heighttrue
                \global\epsfysize=#1
                \ps@dimcent=#1
                \edef\@p@sheight{\number\ps@dimcent}%
}%
\def\@p@@swidth#1{%
                \@widthtrue
                \global\epsfxsize=#1
                \ps@dimcent=#1
                \edef\@p@swidth{\number\ps@dimcent}% 
}%
\def\@p@@srheight#1{%
                \@rheighttrue\use@psfigtrue
                \ps@dimcent=#1
                \edef\@p@srheight{\number\ps@dimcent}%
}%
\def\@p@@srwidth#1{%
                \@rwidthtrue\use@psfigtrue
                \ps@dimcent=#1
                \edef\@p@srwidth{\number\ps@dimcent}%
}%
\def\@p@@sangle#1{%
                \use@psfigtrue
                \@angletrue
                \edef\@p@sangle{#1}%
}%
\def\@p@@ssilent#1{%
                \@verbosefalse
}%
\def\@p@@snoisy#1{%
                \@verbosetrue
}%
\def\@cs@name#1{\csname #1\endcsname}%
\def\@setparms#1=#2,{\@cs@name{@p@@s#1}{#2}}%
\def\ps@init@parms{%
                \@bbllxfalse \@bbllyfalse
                \@bburxfalse \@bburyfalse
                \@heightfalse \@widthfalse
                \@rheightfalse \@rwidthfalse
                \def\@p@sbbllx{}\def\@p@sbblly{}%
                \def\@p@sbburx{}\def\@p@sbbury{}%
                \def\@p@sheight{}\def\@p@swidth{}%
                \def\@p@srheight{}\def\@p@srwidth{}%
                \def\@p@sangle{0}%
                \def\@p@sfile{}%
                \use@psfigfalse
                \@prologfilefalse
                \def\@sc{}%
                \if@noisy
                        \@verbosetrue
                \else
                        \@verbosefalse
                \fi
                \@clipfalse
}%
\def\parse@ps@parms#1{%
                \@psdo\@psfiga:=#1\do
                   {\expandafter\@setparms\@psfiga,}%
\if@prologfile
\fi
}%
\def\bb@missing{%
        \if@verbose
            \typeout{psfig: searching \@p@sfile \space  for bounding box}%
        \fi
        \epsfgetbb{\@p@sfile}%
        \ifepsfbbfound
            \ps@dimcent=\epsfllx bp\edef\@p@sbbllx{\number\ps@dimcent}%
            \ps@dimcent=\epsflly bp\edef\@p@sbblly{\number\ps@dimcent}%
            \ps@dimcent=\epsfurx bp\edef\@p@sbburx{\number\ps@dimcent}%
            \ps@dimcent=\epsfury bp\edef\@p@sbbury{\number\ps@dimcent}%
        \else
            \epsfbbfoundfalse
        \fi
}
\newdimen\p@intvaluex
\newdimen\p@intvaluey
\def\rotate@#1#2{{\dimen0=#1 sp\dimen1=#2 sp
                  \global\p@intvaluex=\cosine\dimen0
                  \dimen3=\sine\dimen1
                  \global\advance\p@intvaluex by -\dimen3
                  \global\p@intvaluey=\sine\dimen0
                  \dimen3=\cosine\dimen1
                  \global\advance\p@intvaluey by \dimen3
                  }}%
\def\compute@bb{%
                \epsfbbfoundfalse
                \if@bbllx\epsfbbfoundtrue\fi
                \if@bblly\epsfbbfoundtrue\fi
                \if@bburx\epsfbbfoundtrue\fi
                \if@bbury\epsfbbfoundtrue\fi
                \ifepsfbbfound\else\bb@missing\fi
                \ifepsfbbfound\else
                \@latexerr{ERROR. cannot locate BoundingBox}\@whattodobb
                        \@p@@sbbllx{100bp}%
                        \@p@@sbblly{100bp}%
                        \@p@@sbburx{200bp}%
                        \@p@@sbbury{200bp}%
                        \no@bbtrue
                        \psdraft
                \fi
                \count203=\@p@sbburx
                \count204=\@p@sbbury
                \advance\count203 by -\@p@sbbllx
                \advance\count204 by -\@p@sbblly
                \edef\ps@bbw{\number\count203}%
                \edef\ps@bbh{\number\count204}%
                 \edef\@bbw{\number\count203}%
                \edef\@bbh{\number\count204}%
               \if@angle
                        \Sine{\@p@sangle}\Cosine{\@p@sangle}%
 
{\ps@dimcent=\maxdimen\xdef\r@p@sbbllx{\number\ps@dimcent}%
 
\xdef\r@p@sbblly{\number\ps@dimcent}%
 
\xdef\r@p@sbburx{-\number\ps@dimcent}%
 
\xdef\r@p@sbbury{-\number\ps@dimcent}}%
                        \def\minmaxtest{%
                           \ifnum\number\p@intvaluex<\r@p@sbbllx
                              \xdef\r@p@sbbllx{\number\p@intvaluex}\fi
                           \ifnum\number\p@intvaluex>\r@p@sbburx
                              \xdef\r@p@sbburx{\number\p@intvaluex}\fi
                           \ifnum\number\p@intvaluey<\r@p@sbblly
                              \xdef\r@p@sbblly{\number\p@intvaluey}\fi
                           \ifnum\number\p@intvaluey>\r@p@sbbury
                              \xdef\r@p@sbbury{\number\p@intvaluey}\fi
                           }%
                        \rotate@{\@p@sbbllx}{\@p@sbblly}%
                        \minmaxtest
                        \rotate@{\@p@sbbllx}{\@p@sbbury}%
                        \minmaxtest
                        \rotate@{\@p@sbburx}{\@p@sbblly}%
                        \minmaxtest
                        \rotate@{\@p@sbburx}{\@p@sbbury}%
                        \minmaxtest
 
\edef\@p@sbbllx{\r@p@sbbllx}\edef\@p@sbblly{\r@p@sbblly}%
 
\edef\@p@sbburx{\r@p@sbburx}\edef\@p@sbbury{\r@p@sbbury}%
                \fi
                \count203=\@p@sbburx
                \count204=\@p@sbbury
                \advance\count203 by -\@p@sbbllx
                \advance\count204 by -\@p@sbblly
                \edef\@bbw{\number\count203}%
                \edef\@bbh{\number\count204}%
}%
\def\in@hundreds#1#2#3{\count240=#2 \count241=#3
                     \count100=\count240        % 100 is first digit #2/#3
                     \divide\count100 by \count241
                     \count101=\count100
                     \multiply\count101 by \count241
                     \advance\count240 by -\count101
                     \multiply\count240 by 10
                     \count101=\count240        %101 is second digit of #2/#3
                     \divide\count101 by \count241
                     \count102=\count101
                     \multiply\count102 by \count241
                     \advance\count240 by -\count102
                     \multiply\count240 by 10
                     \count102=\count240        % 102 is the third digit
                     \divide\count102 by \count241
                     \count200=#1\count205=0
                     \count201=\count200
                        \multiply\count201 by \count100
                        \advance\count205 by \count201
                     \count201=\count200
                        \divide\count201 by 10
                        \multiply\count201 by \count101
                        \advance\count205 by \count201
                     \count201=\count200
                        \divide\count201 by 100
                        \multiply\count201 by \count102
                        \advance\count205 by \count201
                     \edef\@result{\number\count205}%
}%
\def\compute@wfromh{%
                \in@hundreds{\@p@sheight}{\@bbw}{\@bbh}%
                \edef\@p@swidth{\@result}%
}%
\def\compute@hfromw{%
                \in@hundreds{\@p@swidth}{\@bbh}{\@bbw}%
                \edef\@p@sheight{\@result}%
}%
\def\compute@handw{%
                \if@height
                        \if@width
                        \else
                                \compute@wfromh
                        \fi
                \else
                        \if@width
                                \compute@hfromw
                        \else
                                \edef\@p@sheight{\@bbh}%
                                \edef\@p@swidth{\@bbw}%
                        \fi
                \fi
}%
\def\compute@resv{%
                \if@rheight \else \edef\@p@srheight{\@p@sheight} \fi
                \if@rwidth \else \edef\@p@srwidth{\@p@swidth} \fi
}%
\def\compute@sizes{%
        \if@scalefirst\if@angle
        \if@width
           \in@hundreds{\@p@swidth}{\@bbw}{\ps@bbw}%
           \edef\@p@swidth{\@result}%
        \fi
        \if@height
           \in@hundreds{\@p@sheight}{\@bbh}{\ps@bbh}%
           \edef\@p@sheight{\@result}%
        \fi
        \fi\fi
        \compute@handw
        \compute@resv
}
\long\def\graphic@verb#1{\def\next{#1}%
  {\expandafter\graphic@strip\meaning\next}}
\def\graphic@strip#1>{}
\def\graphic@zapspace#1{%
  #1\ifx\graphic@zapspace#1\graphic@zapspace%
  \else\expandafter\graphic@zapspace%
  \fi}
\def\psfig#1{%
\edef\@tempa{\graphic@zapspace#1{}}%
\ifvmode\leavevmode\fi\vbox {%
        \ps@init@parms
        \parse@ps@parms{\@tempa}%
        \ifnum\@psdraft=1
                \typeout{[\@p@sfilefinal]}%
                \if@verbose
                        \typeout{epsfig: using PSFIG macros}%
                \fi
                \psfig@method
        \else
                \epsfig@draft
        \fi
}
}%
\def\graphic@zapspace#1{%
  #1\ifx\graphic@zapspace#1\graphic@zapspace%
  \else\expandafter\graphic@zapspace%
  \fi}
\def\epsfig#1{%
\edef\@tempa{\graphic@zapspace#1{}}%
\ifvmode\leavevmode\fi\vbox {%
        \ps@init@parms
        \parse@ps@parms{\@tempa}%
        \ifnum\@psdraft=1
          \if@angle\use@psfigtrue\fi
          {\ifnum\fig@driver=1\global\use@psfigtrue\fi}%
          {\ifnum\fig@driver=3\global\use@psfigtrue\fi}%
          {\ifnum\fig@driver=4\global\use@psfigtrue\fi}%
          {\ifnum\fig@driver=5\global\use@psfigtrue\fi}%
                \ifuse@psfig
                        \if@verbose
                                \typeout{epsfig: using PSFIG macros}%
                        \fi
                        \psfig@method
                \else
                        \if@verbose
                                \typeout{epsfig: using EPSF macros}%
                        \fi
                        \epsf@method
                \fi
        \else
                \epsfig@draft
        \fi
}%
}%

\def\epsf@method{%
        \epsfbbfoundfalse
        \if@bbllx\epsfbbfoundtrue\fi
        \if@bblly\epsfbbfoundtrue\fi
        \if@bburx\epsfbbfoundtrue\fi
        \if@bbury\epsfbbfoundtrue\fi
        \ifepsfbbfound\else\epsfgetbb{\@p@sfile}\fi
        \ifepsfbbfound
           \typeout{<\@p@sfilefinal>}%
           \epsfig@gofer
        \else
          \@latexerr{ERROR - Cannot locate BoundingBox}\@whattodobb
          \@p@@sbbllx{100bp}%
          \@p@@sbblly{100bp}%
          \@p@@sbburx{200bp}%
          \@p@@sbbury{200bp}%
                \count203=\@p@sbburx
                \count204=\@p@sbbury
                \advance\count203 by -\@p@sbbllx
                \advance\count204 by -\@p@sbblly
                \edef\@bbw{\number\count203}%
                \edef\@bbh{\number\count204}%
          \compute@sizes
          \epsfig@@draft
       \fi
}%
\def\psfig@method{%
        \compute@bb
        \ifepsfbbfound
          \compute@sizes
          \psfig@start
          \vbox to \@p@srheight sp{\hbox to \@p@srwidth 
            sp{\hss}\vss\psfig@end}%
        \else
           \epsfig@draft
        \fi
}%
\def\epsfig@draft{\compute@bb\compute@sizes\epsfig@@draft}%
\def\epsfig@@draft{%
\typeout{<(draft only) \@p@sfilefinal>}%
\if@draftbox
        \hbox{{\fboxsep0pt\fbox{\vbox to \@p@srheight sp{%
        \vss\hbox to \@p@srwidth sp{ \hss 
           \expandafter\Literally\@p@sfilefinal\@nil
                          \hss }\vss
        }}}}%
\else
        \vbox to \@p@srheight sp{%
        \vss\hbox to \@p@srwidth sp{\hss}\vss}%
\fi
}%
\def\Literally#1\@nil{{\tt\graphic@verb{#1}}}
\psfigdriver{dvips}%
\epsfigRestoreAt

\def\pref#1{{(\ref{#1})}}
\def\journal#1&#2(#3){\unskip, \sl #1\ \bf #2 \rm(19#3) }
\def\andjournal#1&#2(#3){\sl #1~\bf #2 \rm (19#3) }

\def\ie{{\it i.e.}}
\def\eg{{\it e.g.}}
\def\cf{{\it c.f.}}

\def\sst{\scriptscriptstyle}

\def\frac#1#2{{#1\over#2}}
\def\coeff#1#2{{\textstyle{#1\over #2}}}
\def\half{\frac12}
\def\hf{{\textstyle\half}}
\def\d{\partial}

\def\inbar{\,\vrule height1.5ex width.4pt depth0pt}
\def\IR{\relax{\rm I\kern-.18em R}}
\def\IC{\relax\hbox{$\inbar\kern-.3em{\rm C}$}}
\def\IH{\relax{\rm I\kern-.18em H}}
\def\IP{\relax{\rm I\kern-.18em P}}

\catcode`\@=11
\def\slash#1{\mathord{\mathpalette\c@ncel{#1}}}
\def\FF{{\cal F}}

\def\NN{{\cal N}}

\def\RR{{\cal R}}
\def\SS{{\cal S}}

\def\underrel#1\over#2{\mathrel{\mathop{\kern\z@#1}\limits_{#2}}}

\catcode`\@=12

\def\vev#1{\left\langle #1 \right\rangle}

\def\tr{{\rm tr}}
\def\Tr{{\rm Tr}}

\def\exp{{\rm exp}}

\def\bps{{\sst BPS}}

\def\eff{{\rm eff}}
\def\aleff{{\alpha'_{\rm eff}}}
\def\sqaleff{{\alpha^{\prime\,1/2}_{\rm eff}}}
\def\aleffsq{{\alpha^{\prime\,2}_{\rm eff}}}
\def\lpl{\ell_{{\rm pl}}}
\def\gym{g_{\sst\rm YM}}
\def\gstr{g_s}

\def\lstr{\ell_s}
\def\Rtil{{\tilde R}}
\def\Ttil{{\tilde T}}
\def\Vtil{{\tilde V}}
\newcommand {\bea}{\begin{eqnarray}}
\newcommand {\eea}{\end{eqnarray}}
\newcommand {\be}{\begin{equation}}
\newcommand {\ee}{\end{equation}}

\let\footnote\savefootnote
\let\footnotetext\savefootnotetext 
\let\footnoterule\savefootnoterule 
\normallatexbib
\begin{document}
%------------ article title  ------------------->>

% For a long title use \\ to cut lines.
% In that case, supply  alternate version of the title
% in square brackets, (it will go in the Table of contents during final
% production of the book.
% \articletitle[Short title]{The long version \\ of this title}

\articletitle[Phases of black holes/branes]{Black Holes and the Phases\\
of Brane Thermodynamics}

%% optional, to supply a shorter version of the title for the running head:
%%\chaptitlerunninghead{}

\author{Emil J. Martinec}

\affil{Enrico Fermi Institute and Department of Physics\\
	University of Chicago\\
	5640 S. Ellis Ave.\\
	Chicago, IL 60637-1433 }         
% Your Institution and address. May cut into  separated 
%                 % lines with \\
%
% optional email address
 \email{e-martinec@uchicago.edu }
%
%%% Repeat the above for multiple authors at different institutions.
%% \author{ }
%% \affil{ }
%% \email{ }

% optional abstract
% \begin{abstract}
% text of the abstract
% \end{abstract}

%------------ body of article ------------------->>
% Write your article here. 
% Note that the \section{section title}
% command allows for the form \section[short title]{very long\\ title}
% Idem for \subsection and \subsubsection

\section{Introduction}

Thermodynamics is a useful tool in analyzing the characteristic
properties of a system of many degrees of freedom.
The equation of state gives the density of states, as well as clues
to the effective degrees of freedom and their interactions,
at the characteristic energy scale set by the temperature.
These lectures will try to convey the idea that brane
thermodynamics is a useful way of organizing much of what
we know about the duality between quantum gravity and
nongravitational theories (often gauge field theories)
which generalizes the AdS/CFT correspondence (see \cite{Aharony:1999ti}
for a review).

We will restrict our considerations to
systems with maximal supersymmetry
compactified on tori; this seemingly simple
arena already encompasses a rich variety of
phenomena.  In section two, the relation
between brane dynamics and black holes is reviewed,
beginning with the correspondence principle \cite{Horowitz:1997nw}
governing the transition between black holes and
objects in perturbative string theory.
Brane subsystems are isolated by the (Maldacena) scaling limit 
\cite{Maldacena:1997re}
that decouples their dynamics from the ambient spacetime;
the UV/IR relation \cite{Susskind:1998dq}
between energy scales in the brane
and gravity descriptions is illustrated in the
context of the thermodynamics.
Criteria are established governing the validity of various
effective descriptions of the thermodynamics
of the system.  In sections three and four, the application of
these criteria are illustrated via a set of increasingly
sophisticated examples.  Section five concludes
with a few closing remarks.  

More details may be found in 
\cite{Martinec:1998ja}\cite{Martinec:1999sa}%
\cite{Martinec:1999gw}\cite{Sahakian:1999gj};
these articles were written
from the perspective of matrix theory \cite{Banks:1997vh}
(see \cite{Banks:1997mn} for a review),
which arises as the universal low-entropy phase of Dp-brane
dynamics on tori.  The present lectures adopt a somewhat
more mainstream viewpoint, adhering more closely to
a description in terms of the gauge theory dynamics
appearing at high entropy.
The phase structure at high entropy, insensitive to
the spatial boundary conditions, was first studied
in \cite{Itzhaki:1998dd},\cite{Barbon:1998cr}.

\section{Branes and black holes}

At weak coupling, the dynamics of $N$ Dp-branes in type II
string theory is governed by open string perturbation theory.
The generic state is a gas of open strings; 
\begin{figure}[ht]
\begin{center}
\epsfig{figure=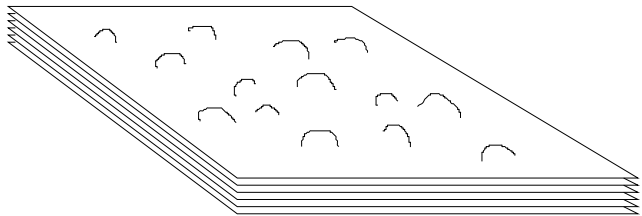,width=.7\textwidth}
\caption{Gas of open strings on a stack of D-branes.}
\label{opengas}
\end{center}
\end{figure}
the low-energy modes
are those of $p+1$-dimensional $U(N)$ supersymmetric
Yang-Mills theory (SYM).  
The low-energy effective action is 
\be
\SS_\eff^{\rm open}=\frac1{\gym^2}
	\int d^{p+1}y\;
	\Tr\left[ F^2+DX^I\cdot DX^I-[X^I,X^J]^2+{\rm fermions}\right]\ ,
\label{Sopen}
\ee
where 
\be
  \gym^2=\gstr\lstr^{p-3}
\ee
is the Yang-Mills coupling on the branes.  For temperatures
$T<\lstr^{-1}$ and weak coupling, the open string gas in a periodic box
of size $\Sigma$ (coordinate identification $y\sim y+\Sigma$
in all spatial directions along the branes) obeys the equation of state
of a perturbative relativistic gas
\bea
  S&\sim& N^2\Sigma^pT^p\nonumber\\
  E&\sim& N^2\Sigma^pT^{p+1}\sim S^{\frac{p+1}{p}}(N^2\Sigma^p)^{-1/p}\ .
\label{gaseos}
\eea
On the other hand, Dp-branes are a source of the gravitational
and other fields of the closed string sector, whose
low energy effective action is
\be
  \SS_\eff^{\rm closed}=\frac1{G_{10}} \int d^{10}x
	\left[e^{-2\phi}\left(\RR +4(\nabla\phi)^2-\coeff1{12}H_{\sst(3)}^2
		\right)-\coeff1{2(p+2)!}\FF_{\sst(p+2)}^2+\ldots\right]\ ,
\label{Sclosed}
\ee
where $\FF_{\sst(p+2)}$ is the RR field strength
sourced by the Dp-branes,
and $G_{10}\sim \gstr^2\lstr^8$.  The generic
state in the superselection sector with $N$ units of RR flux
and energy density concentrated at the source is the
corresponding black Dp-brane geometry.  
\begin{figure}[ht]
\begin{center}
\epsfig{figure=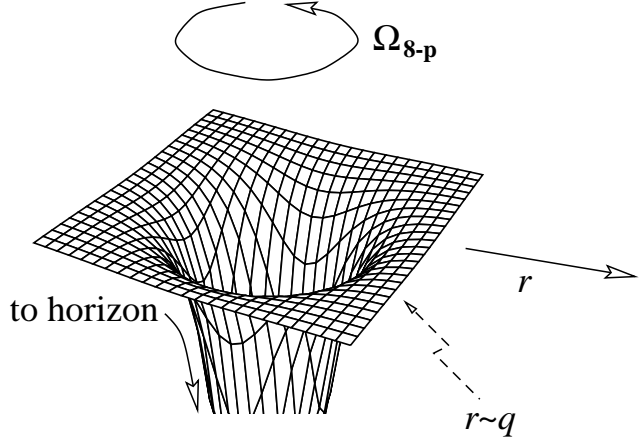,width=.7\textwidth}
\caption{Closed string geometry related to 
the excited D-branes of Figure \ref{opengas},
in the spatial directions transverse to the branes.}
\label{closedgeom}
\end{center}
\end{figure}
The solution of the effective field equations, 
which applies when the source
is macroscopic (\ie\ $N$ is sufficiently large), is
(see \cite{Aharony:1999ti},\cite{Youm:1997hw} for reviews)
\bea
  F_{rty_1...y_p}&=&\d_rH^{-1}\nonumber\\
  e^\phi&=&\gstr H^{\frac{3-p}4}\nonumber\\
  ds^2&=&H^{-1/2}\left(-h\,dt^2+d{\vec y}_{\sst(p)}^2\right)
	+H^{1/2}\left(h^{-1}dr^2+r^2d\Omega_{\sst 8-p}^2\right)\ ,\nonumber\\
  H&=&1+\left(\frac qr\right)^{7-p}\quad,\qquad
  h=1-\left(\frac {r_0}r\right)^{7-p} 
\label{Dpgeom}
\eea
where again the coordinate identification of the directions along
the brane is $y\sim y+\Sigma$.

We might expect to compare the open string gas to closed string geometry
when the system is not too far from extremality,
that is to say not very highly
excited above the (BPS saturated, supersymmetric) 
ground state carrying the given charge:
\be
  E\ll M_{\bps}\Rightarrow r_0\ll q\ ;
\ee
then one has
\bea
  \left(\frac q\lstr\right)^{7-p}&\sim& \gstr N
	\label{rcharge}\\
  \left(\frac {r_0}\lstr\right)^{9-p}&\sim& \frac{S^2}N\gstr^3
	\left(\frac\Sigma\lstr\right)^{-p}\ .
	\label{rhor}
\eea
The first relation is simply Gauss' law for the $(p+1)$-form
RR flux; the second is the usual Bekenstein-Hawking entropy,
determined from the black hole geometry
as the area in Planck units of the horizon at $r=r_0$. 
One can read off the ADM energy of the system from the asymptotic
metric component $g_{tt}\sim -1+(G_{\sst D}M/r^{7-p})$
in the Einstein frame%
\footnote{The Einstein frame is characterized by the field redefinition
$\tilde g_{\mu\nu}= \exp[\alpha\phi]g_{\mu\nu}$
of \pref{Sclosed} that makes the gravitational
term in the effective action 
$G_{\sst D}^{-1}\int\sqrt{\tilde g}\RR[\tilde g]$
in the $D=10-p$ non-compact spacetime dimensions
transverse to the brane.}
\be
  \lstr E\sim\left[\left(\frac{S^2}N\right)^{7-p}
	\left(\frac\Sigma\lstr\right)^{p(p-5)}
	\gstr^{3-p}\right]^{\frac{1}{9-p}}\ .
\label{Dpeos}
\ee
The black geometry has lower free energy than the open string
gas \pref{gaseos} when
(note that $\FF\sim E\sim TS$ in terms of scaling properties)
\be
  \left[\gstr\left(\frac\lstr\Sigma\right)^{p-3}\right]^p =
	\left[\gym^2\Sigma^{3-p}\right]^p
	> S^{3-p}N^{p-6}\ .
\label{corrpt}
\ee
In other words, for large enough coupling $\gym^2$, 
the black geometry takes over.  Defining the effective
coupling at the scale of the temperature $T\sim E/S$
\be
  g_\eff^2=\gym^2NT^{p-3}\ ,
\ee
the condition \pref{corrpt} means that black geometry is
valid for $g_\eff\gg1$, and the open string gas is a
valid description for $g_\eff\ll 1$; an equivalent 
statement is that the geometrical description
holds when the typical radius of curvature of the geometry%
\footnote{One can, for example, use the proper radius 
\pref{Dpgeom} of angular spheres transverse to the branes.
It is straightforward to check that this is string scale
at the point $g_\eff\sim 1$.}
at the horizon satisfies $R_{\rm hor}>\lstr$.
This transition between an ensemble of perturbative strings
and a black geometry is an example of the
{\it correspondence principle} of Horowitz-Polchinski \cite{Horowitz:1997nw},
which states that a black hole becomes a perturbative
string state at sufficiently weak coupling.

One can sharpen the picture considerably by scaling away
the asymptotically flat region, following Maldacena \cite{Maldacena:1997re}:
\be
  \lstr\rightarrow 0 \mbox{~with~}
	\gym^2\Sigma^{3-p},\Sigma E \mbox{~held~fixed}.
\label{decoupling}
\ee
This limit decouples string oscillator modes, whose
energy typically scales as $1/\lstr$.
Since the ten-dimensional gravitational coupling is
$G_{10}\sim \gstr^2\lstr^8$, while the Yang-Mills coupling
on the branes is $\gym^2=\gstr\lstr^{p-3}$, the
closed string dynamics away from the brane source decouples
(\ie\ $G_{10}\rightarrow 0$) for $p<7$
($G_{10}\rightarrow{\it const.}$ for $p=7$).
One also encounters situations described by M-theory;
here, the relevant gravitational coupling is the eleven-dimensional
Newton constant $G_{11}\sim \gstr^3\lstr^9$,
which scales away for $p<6$ and approaches a constant for $p=6$,
in the limit \pref{decoupling}.
Even though the dynamics becomes trivial far from the branes,
one presumes that close enough to the source, the scaling away of the
gravitational coupling is compensated by large
field strengths.  Maldacena conjectured that in
a region of overlap, quantum gravity is defined by
the strongly coupled dynamics of branes; conversely, geometry
gives an effective description of the brane dynamics.

An interesting feature of this relation between geometry
and brane dynamics is a correspondence between increasing distance
from the horizon in the geometry, and high energies
in the brane dynamics -- the {\it UV/IR correspondence}
\cite{Susskind:1998dq}\cite{Peet:1998wn}.
This property is readily seen
in the thermodynamics, where the Hawking temperature $T\sim E/S$ 
of the black geometry is
\be
  T\sim\left(\frac{r_0}{\lstr^2}\right)^{\frac{5-p}2}(\gym^2N)^{-1/2}\ .
\label{THawk}
\ee
Higher temperature (the UV of the brane dynamics) is associated
to larger horizon radius (the IR in gravity), up to $p=5$.%
\footnote{Note that this is the same as the UV/IR relation
obtained in \cite{Peet:1998wn} from the scaling of the wave equation
in the near-horizon geometry.}
The UV/IR correspondence is related to the positive specific
heat of Dp-brane black holes up to $p=5$.  The well-controlled
UV structure of the brane dynamics then provides an IR regulator
for gravity (it implicitly prescribes an asymptotic boundary
condition on the geometry), while the Wilsonian renormalization
group guarantees a well-behaved UV structure for gravity.
This proposal begins to break down for $p=5$, where the
brane dynamics exhibits a limiting temperature
$T\sim(\gym^2N)^{-1/2}$ known as the {\it Hagedorn temperature}.
Such behavior is characteristic of string dynamics, and
so it has been proposed 
\cite{Dijkgraaf:1997hk}\cite{Dijkgraaf:1997nb}\cite{Seiberg:1997zk}
that there is a (strongly coupled)
string theory {\it decoupled from gravity} which governs the
dynamics of fivebranes.  Going a bit further, for $p=6$
the system has negative specific heat -- bulk gravity is not
decoupled, as we saw above; there is no IR regulator, and
arbitrarily many light states couple at longer and longer wavelengths
when we pump energy in to increase the horizon size.
 
To summarize, in regions of strong effective coupling,
the thermodynamics of maximally supersymmetric Yang-Mills theory
is described by/describes black geometries.  The thermodynamics
of black geometry is entirely governed by the geometry at
the horizon; thus we can try to map out a phase diagram
using whatever low-energy description is appropriate
for the horizon geometry in a given strongly-coupled regime,
and matching onto perturbative domains at correspondence points.

To implement this strategy, we should examine the constraints
on the validity of a low-energy geometrical description:
\begin{itemize}
\item
[{\bf (1)}] The curvature at the horizon should be less than the string
scale.  This is the Horowitz-Polchinski correspondence principle.
If this condition is not satisfied, the appropriate description
is a gas of perturbative strings (in the presence of
the background D-branes).
\item
[{\bf (2)}] The string coupling at the horizon should be less than unity.
If this condition is not satisfied, then one either
\begin{itemize}
\item
[{\bf (a)}] lifts to M-theory in a IIA string description; this
repackages the string metric, dilaton, and RR one-form
into an eleven-dimensional metric:
\be
  ds_{11}^2=e^{-2\phi/3}ds_{10}^2
	+e^{4\phi/3}(dx_{11}+C_{\mu}dx^\mu)^2\ .
\ee
D0-branes carry momentum in the eleventh dimension; strings are
membranes wrapped around $x_{11}$; and so on.
\item
[{\bf (b)}] performs an S-duality transformation in a IIB description:
\be
  \phi\rightarrow-\phi\quad,\qquad \gstr\rightarrow 1/\gstr
	\quad,\qquad \lstr^2\rightarrow \gstr\lstr^2 
	\quad (G_{10}\rightarrow G_{10})\ .
\ee
In particular, perturbative strings are interchanged with D1-branes, 
and D3-branes are invariant due to the S-duality symmetry
of 3+1 $\NN=4$ SYM.
\end{itemize}
\item
[{\bf (3)}] Cycle sizes should be greater than the string scale:
\be
  \Sigma_i\sqrt{G_{ii}(r_0)}>\lstr\ .
\ee
If this condition is not satisfied, one should T-dualize
on the corresponding cycles
\bea
  \Sigma_i&\rightarrow& \lstr^2/\Sigma_i\nonumber\\
  \phi&\rightarrow& \phi-\hf\sum_{i=1}^n\log[G_{ii}] \quad,\qquad 
	\gstr\rightarrow \gstr\prod_{i=1}^n\frac{\lstr}{\Sigma_i}\nonumber\\
  G_{ii}&\rightarrow& 1/G_{ii}\ ,
\eea
where the sums and products are over all affected cycles, and
we have assumed a rectangular torus.  
T-duality \cite{Giveon:1994fu}
exchanges momenta on the affected cycles with
perturbative string winding, and mixes the various Dp-branes
\cite{POLCHTASI}\cite{Obers:1998fb}.
For example, Dp-branes wrapping $T^p$ are exchanged with
D0-branes on the dual torus ${\Ttil}^p$.
\item
[{\bf (4)}] After T-duality, one can have a lower-dimensional brane
on a higher-dimensional torus.  If the horizon size is larger
than the proper size of a particular cycle
\be
  \Sigma_i\sqrt{G_{ii}(r_0)}<r_0\ ,
\label{GLinstability}
\ee
the black hole fills that compact dimension; 
if not, it is entropically favorable for
the horizon to localize on that cycle (see Figure \ref{gltrans}).
This instability to topology change of the horizon was
first studied by Gregory and Laflamme \cite{Gregory:1993vy}.
\begin{figure}[ht]
\begin{center}
\epsfig{figure=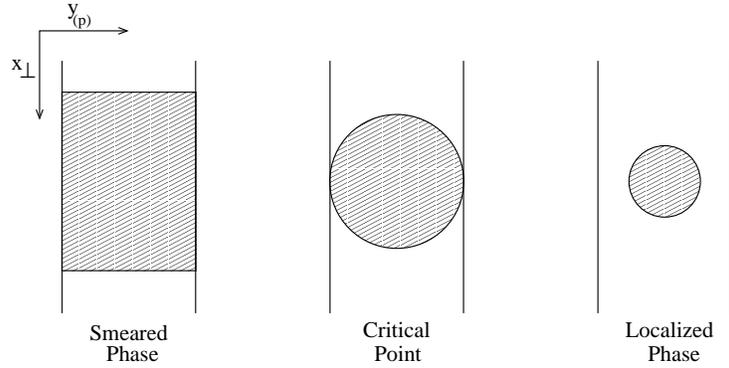,width=.8\textwidth}
\caption{Gregory-Laflamme localization transition}
\label{gltrans}
\end{center}
\end{figure}
\item
[{\bf (5)}] A similar phenomenon occurs in the eleventh dimension,
for D0-brane black holes (see Figure \ref{boostedgltrans}).  
After lifting to M-theory,
D0-brane charge is $p_{11}=N/R_{11}$.  Since
\be
  G_{11,11}=\exp[4\phi/3]\rightarrow\infty \mbox{~as~} r_0\rightarrow 0\ ,
\ee
as the horizon radius decreases, at some point the
horizon localizes on the M-circle parametrized by $x_{11}$.
\begin{figure}[ht]
\begin{center}
\epsfig{figure=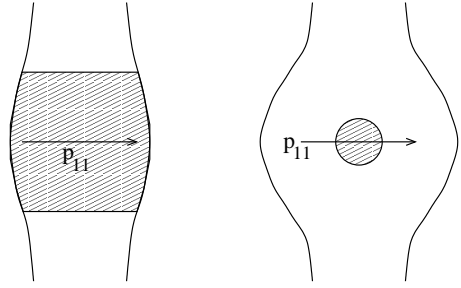,width=.5\textwidth}
\caption{Gregory-Laflamme localization transition for a
boosted horizon}
\label{boostedgltrans}
\end{center}
\end{figure}
The harmonic functions $H$, $h$ in the metric pass from
being independent of $x_{11}$ to being periodic in $x_{11}$.
\end{itemize}

\noindent
To the level of approximation considered here,
all equations of state are scaling relations of the form
\be
  \Sigma E\sim {\rm const.}\times S^aN^bV^{-3c}\ ,
\label{scaling}
\ee
where 
\be
  V^{-3}\equiv \gym^2\Sigma^{3-p}
\ee
is the effective coupling.%
\footnote{The choice of $V$ as the inverse one-third power
of the dimensionless gauge coupling is motivated by the
fact that $V$ is the size of the dual torus $\Ttil^p$
(in 11d Planck units)
seen by D0-branes in low-entropy phases.}
Furthermore, the free energy $\FF$, energy $E$, and $TS$
all scale the same way.
Thus, to find the boundaries between different phases,
we may equate the dimensionless energies 
$(\Sigma E)_{\sst\rm phase 1}=(\Sigma E)_{\sst\rm phase 2}$;
then the phase boundary occurs at
\be
  \frac{\log\,S}{\log\,N}=\alpha\;\frac{\log\,V}{\log\,N}
\ee
in the large N limit.
Consequently, it is convenient to plot the phase diagram as a function of 
$(\frac{\log\,S}{\log\,N})$ versus $(\frac{\log\,V}{\log\,N})$.

Some remarks are appropriate at this point on what is meant
in these lectures by the terms `phase' and `phase transition'.
For the present purpose, we distinguish thermodynamic
phases by the scaling of their equations of state.
At finite brane charge $N$ and in finite volume,
there cannot be any true phase transitions, since the number
of available degrees of freedom will be finite;
rather one has a crossover between limiting behaviors.
In some cases, the transition is a true phase
transition in the classical limit $N\rightarrow\infty$.
An example is the Gregory-Laflamme transition discussed above; 
in \cite{Gregory:1993vy} it was found 
that the classical equations of motion
develop an unstable mode as one enters the regime
\pref{GLinstability}, the hallmark of a second-order
phase transition.
In other examples, such as the Horowitz-Polchinski
correspondence transition, it is not known whether
the classical limit develops a singularity; classical black holes
might or might not smoothly evolve into perturbative classical string
states.  The dynamics of such transitions has been studied in 
\cite{Horowitz:1998jc}\cite{Li:1998jy}; the details
are dimension-dependent.

\section{Dp-brane examples}

\subsection{Warm-up exercise: DO-branes}

\begin{figure}[ht]
\begin{center}
\epsfig{figure=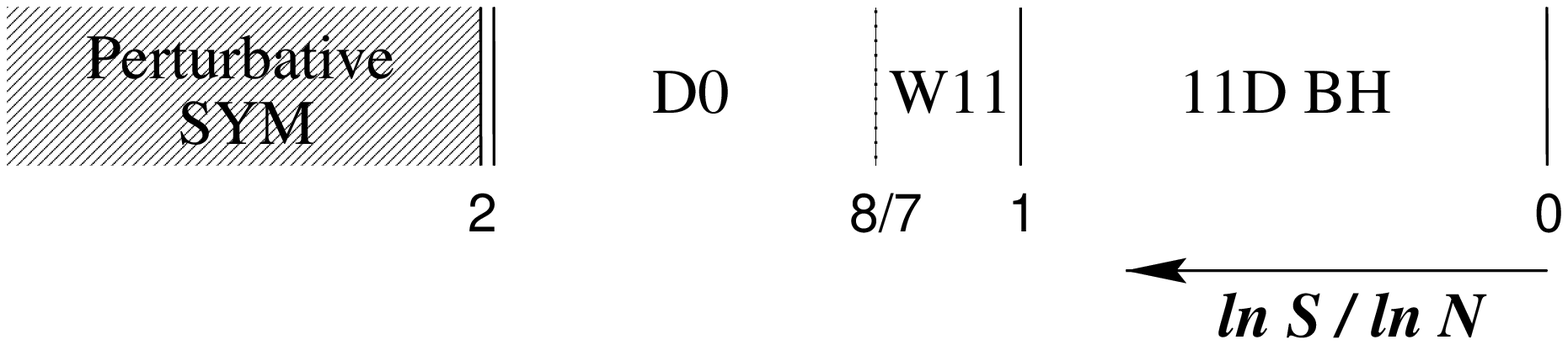,width=.7\textwidth}
\caption{D0-brane phase diagram.}
\label{D0phase}
\end{center}
\end{figure}

Consider first the example of D0-branes.  Here there is
no compactification scale $\Sigma$ since all spatial dimensions
are noncompact (apart from the M-circle).  The Yang-Mills
coupling $\gym$ sets the only scale; and the phase plot
is one-dimensional, since phase boundaries are set by
a condition of the form $(\frac{\log\,S}{\log\,N})=\alpha$.
From \pref{corrpt}, the Horowitz-Polchinski correspondence point 
for $p=0$ is $S\sim N^2$, at which $r_0/\lpl\sim N^{1/3}$.%
\footnote{Here and below, unless otherwise stated,
the Planck scale will always
refer to that of eleven-dimensional M-theory,
$G_{11}\sim \lpl^9\sim \gstr^3\lstr^9$.}  
For $S>N^2$, the horizon radius in Planck units is larger 
than $N^{1/3}$; and from \pref{THawk} we see that 
the Hawking temperature is more than
the energy $r_0/\lstr^2$ required to stretch a string
across the horizon scale $r_0$, indicating that the 
D0-brane quantum mechanics is deconfined.
On the other hand, for $S<N^2$ the thermodynamics
of the ensemble of D0-branes is well-described
by the black D0 geometry.
Strong string coupling $e^\phi|_{\rm hor}\sim 1$
sets in at $S\sim N^{8/7}$, as one sees from 
\pref{Dpgeom},\pref{rhor}.
For lower entropies/energies, the proper effective
description of the horizon geometry (and therefore the
thermodynamics) takes place in eleven dimensional M-theory.
Finally, horizon localization on the M-circle $x_{11}$
occurs for entropies below $S\sim N$.%
\footnote{A rough argument \cite{Banks:1998hz}
is that $p_{11}/M\sim (r_0/S)/(N/R_{11})$
for highly boosted black holes; therefore, for $N\sim S$,
the `Lorentz contraction' $p_{11}/M\sim r_0/R_{11}$ is such
that the horizon is just contained within the M-circle.
Closer inspection \cite{Horowitz:1998fr}
reveals that the black hole horizon
does not Lorentz contract, being a surface of infinite redshift;
rather, the metric backreacts on the momentum stress-energy of the
black hole, expanding by exactly the same factor $p_{11}/M$
(hence the bulge depicted in Figure \ref{boostedgltrans}).  Thus the
net effect is the same: The phase boundary of horizon localization
along the M-circle always occurs at $N\sim S$.}
We can combine all these considerations into the 
phase diagram shown in Figure \ref{D0phase}.
%check location???

\subsection{Dp-branes on $T^p$ for $p>0$}

\begin{figure}[ht]
\begin{center}
\epsfig{figure=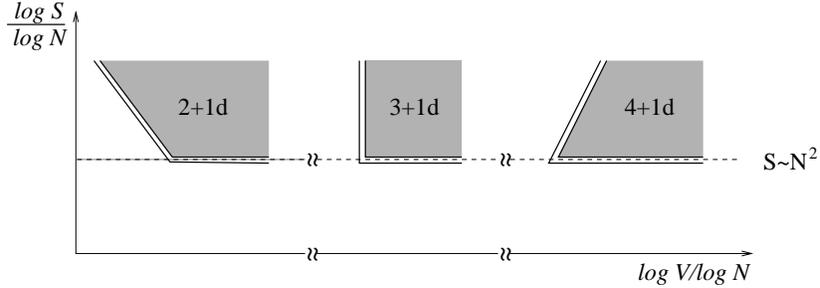,width=.9\textwidth}
\caption{Correspondence transition for various $p$.}
\label{corresptrans}
\end{center}
\end{figure}

Dp-brane thermodynamics for $p>0$ has a phase structure that 
depends on both $\it log\,S/log\,N$ and the dimensionless
Yang-Mills coupling $\it log\,V/log\,N$.  At sufficiently
weak coupling (large $\it log\,V/log\,N$), the phase structure 
for all $p$ is essentially the same as for D0-branes at low enough
entropy, since the dynamics must eventually settle into the
quantum mechanics of the momentum zero-modes on the torus.
Conversely, at high entropy (high temperature), the finite
volume effects are irrelevant; one should see a correspondence
transition as the coupling is increased (smaller $V$).
The correspondence transition is thus bounded by 
Equation \pref{corrpt} or $S\sim N^2$, whichever is met first
coming from weak coupling.  Figure \ref{corresptrans}
illustrates what happens as the dimension increases.  
The shaded region denotes the perturbative gas phase; 
the double line is the correspondence transition curve.  
The horizontal part of the correspondence boundary separates
the perturbative gas from the black D0-brane phase discussed
above.  The angled continuation of the correspondence
curve delineates the transition to the black Dp-brane phase.
The slope of this part of the correspondence curve 
separating the phase of black Dp-branes from the perturbative gas
phase changes with $p$.  The direction of the change is in accord with 
the scaling of the effective coupling with energy (temperature): 
For $p<3$, one passes from the
perturbative phase to the strong coupling, geometrical phase
by decreasing the temperature/entropy; $p=3$ is marginal;
and for $p>3$, strong coupling is met upon increasing the 
temperature/entropy.

One expects that the freezing of the brane dynamics into 
the quantum thermodynamics of its
zero-modes should persist at strong coupling.
The strong coupling thermodynamics is coded in the geometry
of the horizon; the transition to a zero-mode description
is seen as the necessity to perform a T-duality transformation
of the horizon geometry in order to maintain a valid 
effective description, according to criterion (3) of
section 2.%
\footnote{As far as the gauge theory dynamics
is concerned, there is no important distinction between
the T-dual description in terms of
the  quantum mechanics of D0-branes, as opposed to
the dimensional reduction of the Dp-brane gauge
theory, at temperatures sufficiently smaller than $\Sigma^{-1}$.} 
\begin{figure}[ht]
\begin{center}
\epsfig{figure=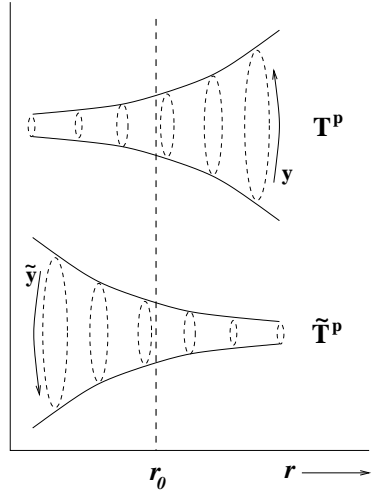,width=.4\textwidth}
\caption{Radial dependence of torus proper size,
as seen by Dp-branes on $T^p$ (top), and D0-branes
on $\Ttil^p$ (bottom).}
\label{dualtorus}
\end{center}
\end{figure}
On the geometrical side, this is seen from the 
proper size of the torus $T^p$ wrapped by the Dp-branes
as the horizon radius decreases.  Below some value of $r$, the
T-dual description, for which the size of $\Ttil^p$ 
increases with decreasing $r$, is the valid one;
see Figure \ref{dualtorus}.
Let us make this change of description;
we are now in a D0-brane geometry at the horizon, smeared over the
dual torus $\Ttil^p$.  As we continue to lower the entropy,
the horizon radius decreases, so eventually we should 
violate the criterion (4), and the horizon localizes on the dual torus.
Also the dilaton runs toward strong coupling at small radius 
in a D0-brane background (Equation \pref{Dpgeom}); at some
point we will also have to pass to an M-theory description.

\subsubsection{D3-branes}
\begin{figure}[ht]
\begin{center}
\epsfig{figure=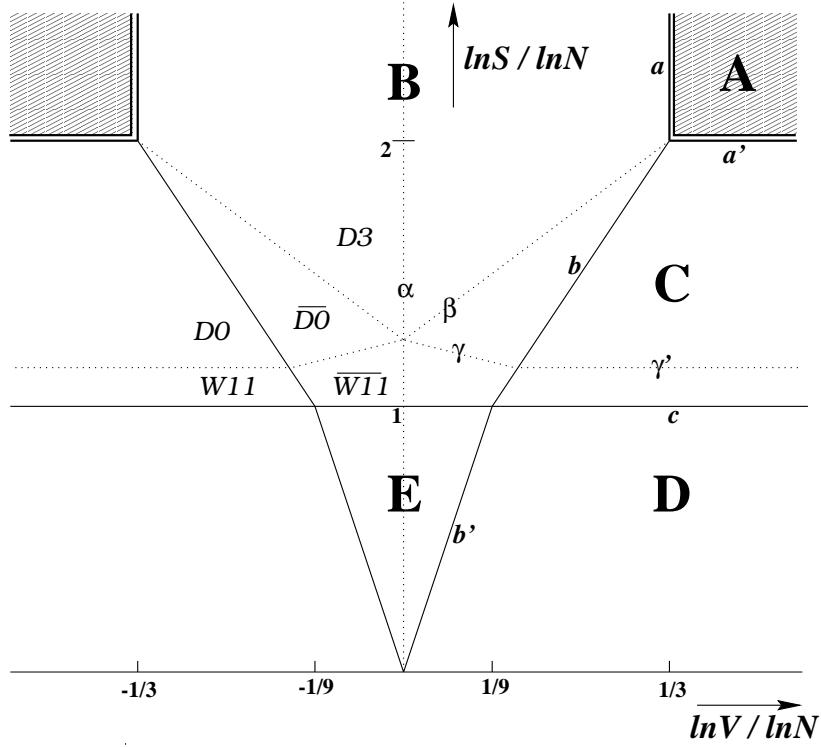,width=.9\textwidth}
\caption{Phase diagram for D3-branes on $T^3$.  
The phases describing various regimes are
{\bf A}: Perturbative D3-brane gas;
{\bf B}: Black D3-brane;
{\bf C}: Black D0-brane;
{\bf D}: Boosted 11d Schwarzschild black hole, localized on $\Ttil^3$;
and {\bf E}: Boosted 11d Schwarzschild black hole, smeared on $\Ttil^3$.
The labels of various patches 
($D0$, $\overline D0$, $W11$, etc.)
and transition curves $a,b,c$; $\alpha,\beta,\gamma$ are 
explained in the text.
The symmetry about the center of the diagram is
a reflection of the S-duality symmetry of 3+1d $\NN=4$ SYM.
}
\label{D3phase}
\end{center}
\end{figure}

Putting together all these considerations leads to a phase diagram
for $V>1$.  This is illustrated in Figure \ref{D3phase}
for the D3-brane case,
%check the location???
where we have extended the structure to $V<1$ using the S-duality
symmetry of maximally supersymmetric 3+1d 
superYang-Mills, which in
the present notation sends $V\rightarrow 1/V$, and reflects
all phase structure about the center of the diagram.
The following phases occur:

\begin{itemize}
\item
{\bf A}: Perturbative D3-brane gas;
\item
{\vskip -\parsep}
{\bf B}: Black D3-brane;
\item
{\vskip -\parsep}
{\bf C}: Black D0-brane;
\item
{\vskip -\parsep}
{\bf D}: Boosted 11d Schwarzschild black hole, localized on $\Ttil^3$;
\item
{\vskip -\parsep}
{\bf E}: Boosted 11d Schwarzschild black hole, smeared on $\Ttil^3$.
\end{itemize}

\noindent
The corresponding regions obtained by S-duality are
labelled similarly.
Phase boundaries are denoted by solid lines, and labelled
by lower case letters: 

\begin{itemize}
\item
{\bf a}: D3-perturbative gas correspondence transition.
\item
{\vskip -\parsep}
{\bf a'}: D0-perturbative gas correspondence transition.
\item
{\vskip -\parsep}
{\bf b}: Gregory-Laflamme horizon localization transition
of the D0-brane horizon on the dual torus $\Ttil^3$.
The horizon localized over
the compact space is denoted $D0$, while
the horizon smeared over the compact space
denoted by an overline $\overline {D0}$;
similarly, $W11$ and $\overline{W11}$ are the lifts
to M-theory of the respective geometries.
\item
{\vskip -\parsep}
{\bf b'}: Continuation of the localization transition
below $S\sim N$ (the slope changes when the phase boundary {\it (c)}
is crossed, due to the change in horizon topology).
\item
{\vskip -\parsep}
{\bf c}: Localization of the horizon on $x_{11}$ at $S\sim N$.
\end{itemize}

\noindent
We distinguish correspondence transitions from localization transitions
by denoting the former with double lines, the latter
by single lines.  Again, to avoid clutter only half of the transition
curves are designated, the others being copies under S-duality.

Within a given phase it may be necessary to perform
duality transformations, lifting to M-theory, or reduction
from M-theory to string theory, in order to maintain a proper
low-energy description of the horizon geometry according
to the criteria of Section 2.
These transformations are changes of description of the
horizon geometry which do not affect the equation of state,
thus are not phase boundaries.  They are denoted by
dotted lines and labelled by greek letters:

\begin{itemize}
\item
$\bf\alpha$: S-duality at $\gym=1$;
\item
{\vskip -\parsep}
$\bf\beta$: T-duality when the proper size of $T^3$ is string scale;
\item
{\vskip -\parsep}
$\bf\gamma$-$\bf\gamma'$: Lift to M-theory when the string coupling
becomes order one at the horizon.
\end{itemize}

\noindent
The patches where different low-energy descriptions hold
within a given phase are indicated on the left half of
Figure \ref{D3phase}.

Note that for $(\frac{\log\,V}{\log\,N})>1/3$ the structure is that of
D0-branes, see Figure \ref{D0phase}.  However, there is a slight difference;
approaching the correspondence curve from above, the 
temperature of the perturbative 
open string gas is $T_+\rightarrow \Sigma^{-1}$;
approaching from below in the geometrical black D0 brane phase,
one has $T_-\rightarrow \Sigma^{-1}(\gym^2N)^{1/3}$, which is
less than $T_+$ all the way until the end of the correspondence
curve at $V^{-1}=(\gym^2N)^{1/3}\sim 1$.  This indicates
an intermediate phase, not visible when the phase structure is
plotted as a function of entropy.  One imagines that as this
isentropic phase is traversed, the holonomy
of the U(N) gauge fields on the torus become frozen, 
since they are fluid for high entropy 
and frozen \cite{Banks:1998hz} for low entropy.

\subsubsection{D2-branes and D4-branes}
\begin{figure}[ht]
\begin{center}
\epsfig{figure=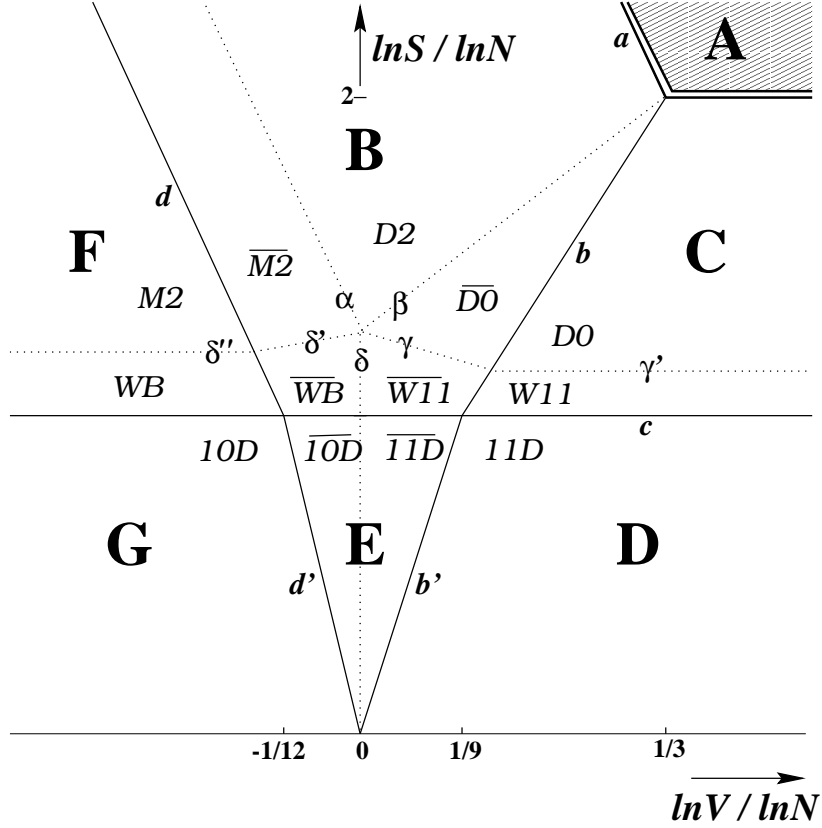,width=.9\textwidth}
\caption{D2-brane phase diagram.  The phases are --
{\bf A}:~Perturbative D2-brane gas.
{\bf B}:~Black D2-brane.
{\bf C}:~Black D0-brane.
{\bf D}:~Boosted 11d Schwarzschild black hole, localized on $\Ttil^2$.
{\bf E}:~Boosted 11d Schwarzschild black hole, smeared on $\Ttil^2$.
{\bf F}:~Black M2-brane.
{\bf G}:~Boosted 10d Schwarzshild black hole, localized on ${\tilde S}^1$.
}
\label{D2phase}
\end{center}
\end{figure}

\begin{figure}[ht]
\begin{center}
\epsfig{figure=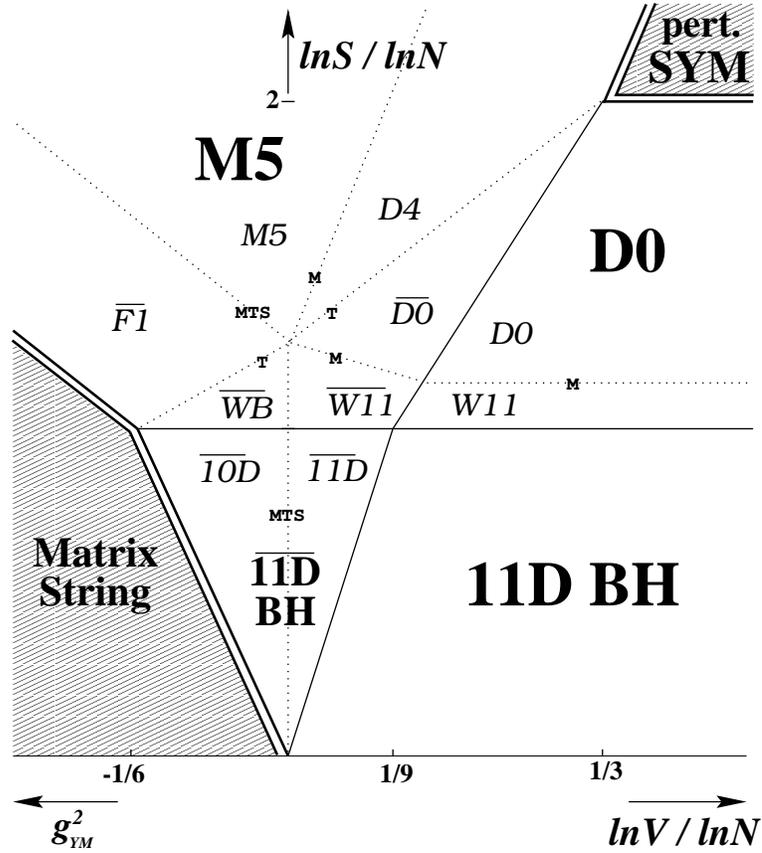,width=.9\textwidth}
\caption{D4-brane phase diagram.}
\label{D4phase}
\end{center}
\end{figure}

The structure of the phase diagram for $V>1$ is essentially
the same for all Dp-branes wrapping $T^p$, differing only
in the slope of the corresponence curve, Figure \ref{corresptrans}.
The analogous phases and transition curves carry the
same labels as in Figure \ref{D3phase}. 
However, for $p\ne3$, there is typically no symmetry
relating the phase diagram at $V<1$ to that for $V>1$, 
and further analysis is necessary for this regime.  
The details depend on $p$, and a case-by-case 
analysis is necessary.  
We will illustrate with the examples of D2-branes and D4-branes
(see Figures \ref{D2phase} and \ref{D4phase}).
Small $V$ corresponds to large coupling in the Dp-brane gauge theory,
at the scale of the torus cycle size; thus one can think
of the passage to small $V$ at large entropy/temperature (where 
$p+1$ dimensional kinematics applies) as the strong coupling limit.
For $p=2$, this is a flow to the IR; for $p=4$ one flows to the UV.

Maximally supersymmetric Yang-Mills theory in 2+1d
flows to a nontrivial fixed point in the IR.  On the geometrical
side, one sees this in the thermodynamics as a transition 
of the horizon geometry to a structure that is locally 
$AdS_{4}\times S^7$, with an equation of state
\be
  E\sim S^{3/2}[N^{3/2} \Sigma^2]^{-1/2}\ .
\label{M2eos}
\ee
The energy-versus-entropy scaling is
determined by conformal invariance; the effective number
of field theoretic degrees of freedom is the coefficient
of the spatial volume $V^p$ in the last factor, 
\ie\ $N^{3/2}$;%
\footnote{On the other hand, D3-branes have $N^2$
degrees of freedom -- one can track them from weak coupling
\pref{gaseos} to strong coupling \pref{Dpeos} along a marginal line.}
it would be interesting to have
an explanation of this scaling from the gauge theory side.
The scaling \pref{M2eos} differs from that of the D2-brane,
Equation \pref{Dpeos}.  Starting from the black D2 correspondence
curve and continuing to the left at constant entropy,
one first lifts to the M-theory black M2-brane
geometry when the dilaton at the horizon becomes order one
(the dotted line labelled $\alpha$ in the figure);
the M2-brane being pointlike in the eleventh dimension,
the horizon can localize along the M-circle, and does so
along the transition curve {\it (d)} in Figure \ref{D2phase}
(the curve {\it (d')} is its continuation below $S\sim N$).
It is this localized M2-brane phase that has the
equation of state \pref{M2eos}.  

If we now start in this M2-brane phase, and pass to smaller
entropy/temperature at fixed $V$, the size of the $T^2$
wrapped by the M2-branes (measured at the horizon) shrinks
until it becomes Planck size.  
M-theory on a sub-Planckian two-torus is dual to a
circle in IIB string theory
\cite{Schwarz:1995dk}\cite{Aspinwall:1995fw},
and the M2-brane stack dualizes into a black wave 
along this circle.  This change of description is the
dotted line ($\delta$-$\delta'$-$\delta^{\prime\prime}$) in the figure.
Eventually the wave localizes along this circle to a
boosted black hole in IIB string theory.  All told, the
set of phases is

\begin{tabular}
l\\
{\bf A}: Perturbative D2-brane gas.\\
{\bf B}: Black D2-brane.\\
{\bf C}: Black D0-brane.\\
{\bf D}: Boosted 11d Schwarzschild black hole, localized on $\Ttil^2$.\\
{\bf E}: Boosted 11d Schwarzschild black hole, smeared on $\Ttil^2$.\\
{\bf F}: Black M2-brane.\\
{\bf G}: Boosted 10d Schwarzshild black hole, localized on ${\tilde S}^1$.\\
\end{tabular}
\bigskip

\noindent
Phases {\bf A}-{\bf E} are analogues of those appearing in
the D3-brane case; the strong-coupling phases {\bf F} and {\bf G}
are new.

Maximally supersymmetric Yang-Mills in 4+1d flows to
strong coupling in the UV, and thus one must find the strong coupling
fixed point which it descends from in order that the
theory is well-defined.  Geometry again supplies the answer 
\cite{Rozali:1997cb}\cite{Seiberg:1997ad}\cite{Itzhaki:1998dd}:
The D4-brane geometry lifts to M-theory five-brane at strong coupling,
and the required fixed point is a six-dimensional nonabelian tensor
field theory with (2,0) supersymmetry, about which little is
known.  From \pref{Dpeos},
the equation of state of this six-dimensional theory is
\be
  E\sim S^{6/5}[N^3 (\gym^2 \Sigma^4)]^{-1/5} \ ,
\label{D4eos}
\ee
and one can check that the horizon geometry of the
M-lifted metric \pref{Dpgeom} is locally $AdS^7\times S^4$.
Because the M-circle lies parallel to the brane, in this case there
can be no localization transition of the horizon along this direction;
hence the D4 equation of state merely needs to
be reinterpreted as that of a higher-dimensional field theory.
We see that the theory lives on a torus, four of whose coordinates
have period $\Sigma$, while the fifth is of size $\gym^2$
\cite{Rozali:1997cb};
the effective number of degrees of freedom appears to be $N^3$
(more about this in section 4 below).
%check location???

By now the structure of duality transformations/phase transitions
should be familiar, so they are labelled directly
on Figure \ref{D4phase}.  One new feature -- a correspondence transition
at strong coupling -- does arise at the lower left.
As the entropy/temperature is lowered in this regime, 
The horizon size of the original $T^4$ shrinks, while
the M-circle remains somewhat larger.  Thus at some point
we will want to reduce to IIB string theory along the $T^4$.%
\footnote{The corresponding dotted line on the figure is
labelled MTS, since one way to approach the problem is to
select one of the cycles of the $T^4$ for reduction to IIA
string theory; then one finds that the remaining three
are substringy and must be T-dualized, leading to IIB string
theory; finally, the resulting string coupling is larger
than one, so an S-duality must be performed.  At this
point all the criteria of section 2 are met.}
This sequence of transformations turns the D4/M5-brane charge
into IIB perturbative string winding around the M-circle 
(the circle related to the 4+1d SYM coupling).
The resulting perturbative string should have a correspondence
transition where a thermally excited, perturbative
macroscopic string has less free energy than the black geometry.
Thus the phase labelled `Matrix String' in the figure:
A perturbative string arises from the strong coupling
dynamics of a gauge theory, as in 
\cite{Motl:1997th}\cite{Banks:1997my}\cite{Dijkgraaf:1997vv}.

\subsection{Qualitative checks\label{qualcheck}}

The strong-coupling regime of the phase diagram has been evaluated
using black geometry.  How much of this structure can we
motivate from the gauge theory side?
\begin{itemize}
\item
$S<N$: One can take the D0-brane interactions determined
from the Born-Oppenheimer approximation; the virial theorem
and mean field theory then lead to the equation of state
scaling of boosted 11d black holes 
\cite{Banks:1998hz}\cite{Horowitz:1998fr}\cite{Li:1998ci}
(in the same spirit as one estimates the size and properties
of an atom using the virial theorem and uncertainty principle).
\item
$N<S<N^2$: Miao Li \cite{Li:1999qj} has given the following estimate of
the properties of the localized black D0-brane phase:
Suppose the matrix eigenvalue distribution of the D0-brane
quantum mechanics is given by
\be
  \rho(r)=|\psi(r)|^2\propto\frac{N}{L^9}\left(\frac rL\right)^\alpha\ ,
\ee
with $L=N^{1/3}\lpl$ the scale set by 't Hooft scaling
(see the discussion in section 3.1).
The scaling exponent $\alpha$ is determined by demanding that
the free energy of off-diagonal matrix elements reproduces
the equation of state
\be
  E\sim \frac{R_{11}}{\lpl^2}\left(\frac{S^2}{N}\right)^{7/9}
\ee
in the mean field theory approximation; then 
$\coeff 1N\vev{\tr[X^2]}\sim r_0^2$
correctly reproduces the horizon scale as a 
nontrivial self-consistency check.
\item
$S\sim N^2$ (from below): as discussed in section 3.1,
at this point the Hawking temperature matches the energy
required to excite the off-diagonal matrix elements
of the D0-brane quantum mechanics, which are the stretched
strings between the D0 particles; this indicates
that a deconfinement transition is taking place.
\item
{\it Horizon localization}: The Gregory-Laflamme localization
transition take place uniformly for all $p$ along
a curve $S\sim V^{9/2}N^{1/2}$.  
Susskind \cite{Susskind:1997dr} has argued
that this is an analogue of the Gross-Witten large $N$
phase transition of pure 1+1d gauge theory on a circle.  
In the Gross-Witten toy model, 
the eigenvalues of the gauge field zero-mode 
(the only gauge-invariant degrees of freedom) undergo
a phase transition.  The eigenvalues themselves live on a circle,
and at high temperature they fill the circle, 
while at low temperature they clump together. 
Similarly, the locations of D0-branes on a $p$-torus 
$\Ttil^p$ are the eigenvalues of the gauge field on the dual torus $T^p$.
A smeared horizon has the D0-branes spread across the torus cycles,
while a localized horizon clumps the eigenvalues together.
The details probably differ from the toy model.  For instance,
the Gross-Witten transition is third order, whereas the 
Gregory-Laflamme transition proceeds via the development
of a soft mode in the perturbation spectrum of the horizon,
leading one to expect a second-order transition.
\item
{\it Qualitative phase structure}.  
Determined above via geometry, the phase structure matches
what is known for the strongly-coupled gauge theory.
Maximally supersymmetric Yang-Mills has strong-coupling fixed
points for $p=1,2,3,4$ \cite{Seiberg:1997ax}, 
and we have seen in the examples that
the appropriate regime of the phase diagram has the equation
of state scaling in the manner dictated by conformal invariance.
\end{itemize}

\noindent 
One of the intriguing issues is why such obviously naive
arguments seem to be so successful in capturing
the properties of the strong coupling, geometrical phases;
for instance, why does mean field theory 
yield the correct scaling exponents?

\subsection{Phases of 3+1d SYM on $S^3$}

\begin{figure}[ht]
\begin{center}
\epsfig{figure=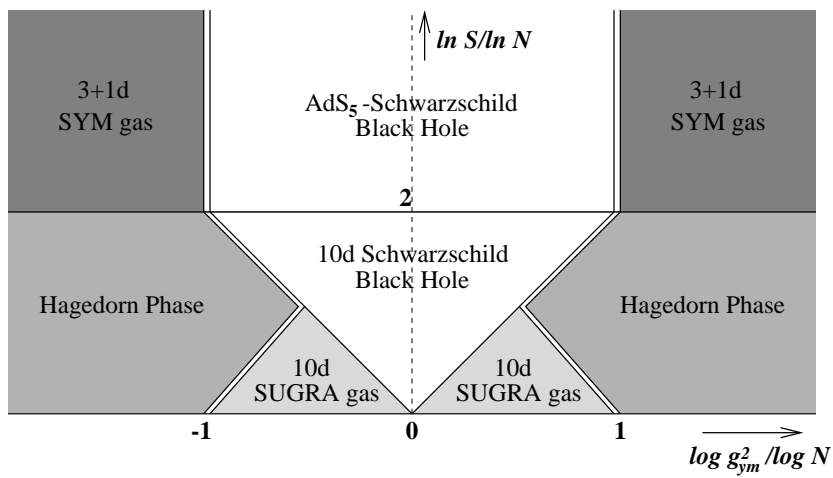,width=.9\textwidth}
\caption{Phase diagram of D3-branes on $S^3$.}
\label{D3sphere}
\end{center}
\end{figure}

The low-entropy/temperature structure of brane dynamics
on a torus depends heavily on the global topology; 
the $E_{p+1}$ U-duality \cite{Obers:1998fb} 
was an essential ingredient
in maintaining a valid effective description of
the horizon geometry in strong effective coupling.
There is no such duality group for gauge theory on
the sphere, so we should expect a rather different
phase structure.  Nevertheless, when geometry describes
the thermodynamics, we expect the same sorts of
phenomena: correspondence transitions bound
the validity of the geometrical description; and
localization transitions can occur when the horizon
is of the same order as a cycle size,
if the brane is not wrapped around that cycle.
The scalings of the various transitions were worked
out in \cite{Banks:1998dd}, and the results are plotted in
Figure \ref{D3sphere}.  
The structure at $S>N^2$ is the same
as for the three-torus, since here the thermal wavelength
is smaller than the size of the compact space and
the boundary conditions are irrelevant.  The perturbative
open string gas has a correspondence transition when the 't~Hooft
coupling (or its dual) is of order one,
to an $AdS_5\times S^5$ Schwarzschild black hole;
the horizon size is larger than (and hence smeared across)
the five-sphere transverse to the branes.
There is a localization transition in this strong-coupling
region at $S\sim N^2$ (horizon radius of order the
five-sphere scale), to a ten-dimensional Schwarzschild
black hole completely localized on $AdS_5\times S^5$.
The free energy of a perturbative string (whose radius of
gyration is much larger than the radius of curvature of
the ambient anti-deSitter space) wins out over 
this black hole at sufficiently weak coupling,
leading to a continuation of the correspondence
curve below $S\sim N^2$.  A dilute gas of gravitons
in $AdS_5\times S^5$ dominates at the lowest entropies
at strong coupling;
the phase transition to a 10d Schwarzschild black hole
is just the gravitational Jeans instability.
Note that both the Hagedorn phase and the localized 10d black hole
phase would be missed in an analysis using the canonical ensemble;
since these two phases do not have positive specific heat,
coupling the system to a heat bath pumps energy/entropy
into the system such that one jumps directly from the supergravity
gas phase to the smeared black hole phase above $S\sim N^2$.

\section{The D1-D5 system and phases of\\ `little string' theory}

As a final illustration of the application of thermodynamic
techniques to learn about the properties of brane dynamics,
consider \cite{Martinec:1999gw} the D1-D5 system compactified on a torus 
$T^4\times S^1$, with the four-torus having volume $V_4$
and the circle having radius $R$.  
This system has a rich history (see for example \cite{Maldacena:1997tm}).
Here, we will see in the thermodynamics some hints of
the origin of the apparently $O(N^3)$ degrees of freedom
of the M-theory fivebrane observed in \pref{D4eos}.

We have seen that, as the entropy is lowered, U-duality
transmutes brane charge into momentum; we obtained the
infinite momentum frame description of Schwarzschild black
holes in M-theory.  Adding a second charge to the system,
we should expect that we will now find a boosted brane
at low entropies.  The D1-D5 system is such that
we find the IMF description of M-theory fivebranes
at low momentum \cite{Aharony:1998th}.
The associated classical geometry is
\bea
  ds^2&=&(H_1H_5)^{-1/2}(-h\,dt^2+dy_5^2)+
	H_1^{1/2}H_5^{-1/2}(dy_1^2+...+dy_4^2)\nonumber\\
	& &\qquad\qquad+(H_1H_5)^{1/2}(h^{-1}dr^2+r^2d\Omega_3^2)\nonumber\\
& &\nonumber\\
  H_a&=&1+\frac{q_a^2}{r^2}\quad a=1,5\quad;\qquad h=1-\frac{r_0^2}{r^2}\ ,
\label{D1D5geom}
\eea
with $y_i\approx y_i+V_4^{1/4}$, $i=1...4$, and $y_5\approx y_5+R$.
One can think of this system as a stack of $Q_5$ fivebranes
in a superselection sector with $Q_1$ embedded one-branes.
A D1-brane inside a D5-brane is the same
as an instanton in the D5 gauge theory 
\cite{Douglas:1995bn}\cite{Douglas:1996uz};
since instantons have codimension four, the instanton
in 5+1 superYang-Mills is a 1+1d soliton (effectively a string).
The instanton moduli space has bosonic dimension of
order $Q_1Q_5$, the number of perturbative 1-5 strings;
thus one might expect that $k\equiv Q_1Q_5$ is the measure of
the number of degrees of freedom in the system.

Without the D1 charge $Q_1$, we saw (Equation \pref{THawk})
that the thermodynamics of $Q_5$ D5-branes had a
limiting temperature -- Hagedorn behavior --
characteristic of an underlying string with an (inverse) tension 
\be
  T_{\rm max}\sim(\aleff)^{-1/2} \quad,\qquad 
  	\aleff=\gym^2 Q_5
\ee
(Note that $\gym^2=\gstr^2\lstr^2$ here, so that
$\aleff\ne\lstr^2$!).  The limit which decouples the asymptotically
flat region is 
\be
  \lstr\rightarrow 0 \mbox{~with~}
	E{\sqaleff}\ ,\quad\frac{R}{\sqaleff}\ ,\quad
	\mbox{and~}\frac{V_4}{\aleffsq}\quad\mbox{held~fixed.}
\ee
Note that this is {\it NOT} the limit usually taken -- 
$\lstr\rightarrow0$, with $ER$, $\gstr$, and $V_4/\lstr^4$ held fixed -- 
which yields the much-studied $AdS_3\times S^3\times T^4$ near-horizon
geometry (\cf\ A. Strominger's lectures).  This latter limit shrinks
away the four-torus transverse to the D1 winding, so that the
system becomes 1+1 dimensional.  Here we are interested in
maintaining the full 5+1 dimensional nature of the little string
dynamics.  In terms of the radial scales in the geometry,
we are taking $q_5\gg q_1,r_0$; different regimes will arise
depending on whether $q_1>r_0$ or $q_1<r_0$.

The energy in the system can again be read off the 5d Einstein frame
metric, and the entropy is as always the horizon area in (10d)
Planck units.  After a bit of algebra, one can process the
equation of state into the following form: At high entropies,
the strongly coupled equation of state is
\be
  E=-p_{||}+\sqrt{(S/2\pi\sqaleff)^2+p_{||}^2}\ ,
\label{hageos}
\ee
where $p_{||}=kR/\aleff\equiv k/\Rtil$  (\ie\ $\Rtil=\aleff/R$)
behaves as an effective momentum carried by the system --
the equation of state is that of a relativistic system
of invariant mass $M=S/2\pi\sqaleff$.  Thus we have a 
boosted Hagedorn gas.  In the infinite momentum frame limit
$p_{||}\gg M$, we recover the equation of state 
$S=2\pi\sqrt{k\cdot 2RE}$ of BTZ black holes; this is the further
limit required to obtain the $AdS_3$ horizon geometry.

Since both types of brane are wrapped around the circle of
radius $R$, this circle shrinks the fastest with decreasing 
horizon radius; we should expect to have to T-dualize it
relatively early on in applying the criteria of section 2.
The resulting D0-D4 system can localize on the dual circle
(loosely speaking, the one of radius $\Rtil$).  The localization
transition is nothing other than the Hagedorn transition
of the effective string!  The equation of state changes to
\bea
  E&\sim& S^{6/5}\left[\frac{Q_5^3\gym^2V_4}{R}\right]^{-1/5}
	\quad,\qquad p_{||}\ll E\ (r_0\gg q_1) 
	\label{m5gas}\\
& &\nonumber\\
  E&\sim& S^3\left[\frac{\gym^2}{k^{3/2}V_4^{1/4}R}\right]
	\quad\qquad,\qquad p_{||}\gg E\ (r_0\ll q_1) \ .
	\label{dlcqmembrane}
\eea
We can understand the first of these in the following way:
$\gym^2/R$ in 5+1d is the 4+1d Yang-Mills coupling,
so \pref{m5gas} is the same equation of state as the fivebrane
gas encountered in section 3.2.2.  Thus, at
relatively low momenta (entropy dominating
over the boost $\sqaleff p_{||}$), the Hagedorn transition
is one of long, thermally excited effective strings overtaking
the gas of modes of the (2,0) supersymmetric tensor
field theory that describes the fivebrane
dynamics at lower energies.

The second equation of state, Equation \pref{dlcqmembrane},
should describe the `zero-mode' dynamics of the 
toroidally compactified fivebrane, if our previous examples
serve as an accurate guide.  In these single-charge brane
systems, we found that the low-entropy structure obeyed
infinite momentum frame kinematics of highly boosted objects;
indeed, the Hagedorn phase \pref{hageos} exhibits 
precisely such behavior.  To cast \pref{dlcqmembrane}
in this form, we posit that the strongly coupled
system occupies an effective transverse four-torus of size
\be
  \Vtil_4=\frac{Q_5}{Q_1}V_4\ ;
\ee
this is reminiscent of other gauge systems 
\cite{Maldacena:1996ds}\cite{Banks:1998hz}
where at low entropies the holonomies of the gauge fields
arrange themselves in such a way as to modify the box size seen
by the effective degrees of freedom.  With this redefinition,
and the apparent longitudinal box size $\Rtil$ arising in
the Hagedorn phase, we can recast the localized equations of
state \pref{m5gas}, \pref{dlcqmembrane} as
\bea
  E&\sim& S^{6/5}\left[k\Vtil_4\Rtil\right]^{-1/5}
        \quad\ ,\qquad p_{||}\ll E 
	\label{newM5gas}\\
& &\nonumber\\
  E&\sim& \frac{1}{p_{||}}
		\left[\frac{S^{3/2}}{(k\Vtil_4^{1/2})^{1/2}}\right]^2
        \quad,\qquad p_{||}\gg E\ .
        \label{newdlcqmembrane}
\eea
Note that these parameter redefinitions uniformly encode the
data in all strong-coupling phases.  
Most importantly, the $Q_5^3$ count of degrees of freedom
in the `bare' equation of state has been processed into
the `dressed' count of $k=Q_1Q_5$ degrees of freedom,
which is more like what we were expecting.
Also, equation of state
\pref{newdlcqmembrane} now has the form required by 
infinite momentum frame kinematics, suggesting an object of
invariant mass $M\sim S^{3/2}/(k\Vtil_4^{1/2})^{1/2}$ replaces
the boosted Hagedorn string below the Hagedorn transtion.
This is the equation of state of a 2+1d relativistic gas
of $k$ degrees of freedom, living in an effective box
of size $\Vtil_4^{1/2}$.  A good candidate remains to
be found for the composition of this effective membrane-like object.
\begin{figure}[ht]
\begin{center}
\epsfig{figure=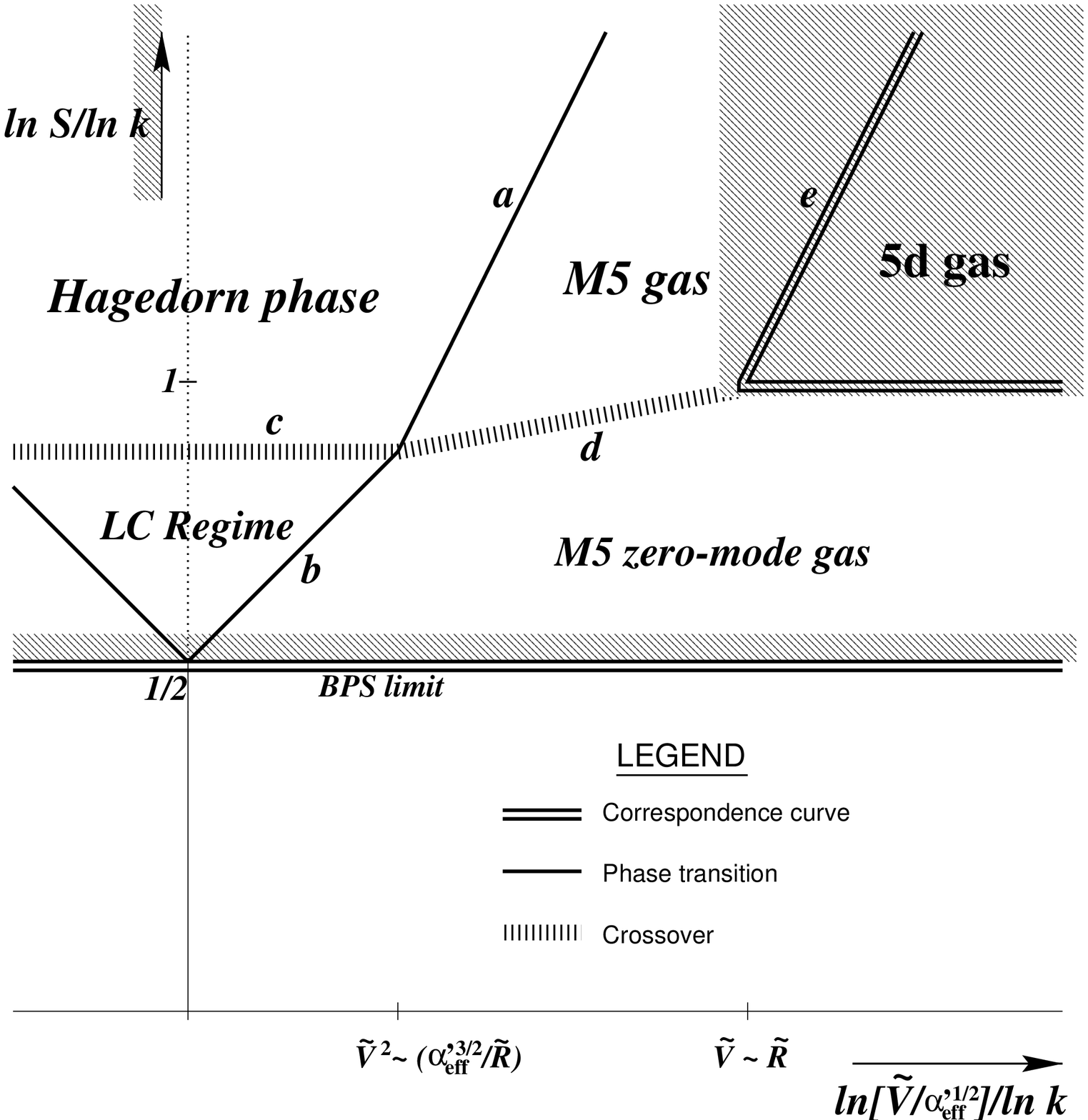,width=.7\textwidth}
\caption{Phase diagram of boosted M5-brane `little strings'.}
\label{D1D5phase}
\end{center}
\end{figure}

The phase diagram is plotted in Figure \ref{D1D5phase}.  
The wide dashed line indicates the location of the crossover 
to infinite momentum frame kinematics at
$p_{||}\sim E$, with the IMF side being that of lower entropy.
Further evidence that the reshuffling of the parameters 
is on the right track comes from the scaling 
of the various transition curves {\bf a}-{\bf e}
\cite{Martinec:1999gw}, which are all of the form
\be
  \frac Sk\sim\left(\frac{\Rtil}{\sqaleff}\right)^\alpha
	\left(\frac{\Vtil_4}{\aleffsq}\right)^\beta\ ,
\ee
\ie\ the entropy per degree of freedom at the transition
is determined entirely by the effective geometry;
the expressions in terms of the bare parameters has
no such uniform interpretation.

Remarkably, a system that we otherwise understand very poorly,
being some sort of strongly coupled string theory without gravity,
has its thermodynamics controlled by black geometry!
In particular, the Hagedorn transition of the effective
string is a geometrical transition of horizon localization.
As in the matrix theory of 11d supergravity, a system
whose underlying degrees of freedom are otherwise
hard to understand, the introduction of a superselection charge
with the interpretation of momentum leads to a simplification.
In fact, the analogy is rather direct: The matrix model
of 11d supergravity describes a system with longitudinal
momentum $p_{||}=N/R_{11}$, in terms of the $N\times N$ matrix
quantum mechanics of $N$ D0-branes; here, the matrix model of M5-branes
appears to carry boost $p_{||}=k/\Rtil$ and have $k$ degrees
of freedom, which might be traced back to the dynamics of
the $Q_1\times Q_5$ matrix of 1-5 strings.

Much remains to be understood, however.  The forms of the equation of 
state \pref{newdlcqmembrane},\pref{newM5gas} are the low- and high-entropy
asymptotics of the full equation of state derived from
the D0-D4 black hole localized on a transverse circle,
in the limits $r_0\ll q_1$ and $r_0\gg q_1$,
respectively.  One would like to understand the finite size
effects that give rise to the membrane-like structure
apparent in \pref{newdlcqmembrane}; these do not have the
structure of a canonical boosting 
as in \pref{hageos} \cite{Martinec:1999gw}.%
\footnote{On the other hand, when the boost is along a direction
in which the horizon is smeared, generically the equation of
state does {\it not} have a canonically boosted form;
so we might turn the problem around, and ask why the
Hagedorn equation of state \pref{hageos} {\it does}
have a canonically boosted form.  Nevertheless the low-entropy
limit of both the Hagedorn equation of state and 
\pref{newdlcqmembrane} has the
standard infinite momentum frame form.}
It would be helpful to have a model to explain how the effective
box sizes $\Rtil$ and $\Vtil_4$ arise dynamically.
For instance, $\Rtil$ was motivated as the `T-dual'
of the original winding radius $R$ of the D1-branes,
using the effective string scale $\aleff$; however, the little
string theory does not really have such a global duality symmetry.
The factor of $Q_5$ in $\Vtil_4$ is likely the result of global
holonomy of the D5 gauge fields linking together the individual
fivebranes as one traverses the cycle of the four-torus; the
inverse dependence on $Q_1$ seems puzzling, though.

\section{Concluding remarks}

It is remarkable that such a wide range of parameter space
in the thermodynamics of maximally supersymmetric gauge theory
is accessible to study via a set of dual geometrical
descriptions.  
Moreover, the dual geometrical descriptions exemplify
in a unifying fashion much of the duality structure
of M-theory on tori, linking together the Maldacena
and matrix conjectures.  Thermodynamic considerations
are one of the few non-BPS probes we have of
strongly couped brane systems, and may (as in section 4)
lead us to new insights about their effective
degrees of freedom at different scales.

The brane ensemble at finite N 
undergoes decay via Hawking radiation; thus the time
evolution of the system gradually traces a path on the phase diagram.
In our analysis, we have ignored the fact that
the brane dynamics on a torus has a large moduli space
of flat directions, corresponding to separating the
branes transversely; in fact this is perhaps the simplest
decay mode for very near-extremal black holes 
composed out of toroidally wrapped branes.  
Thermodynamics is valid in the large
N limit, when the decay time can be made arbitrarily long
relative to other time scales in the dynamics.
But it would seem that we can probe
arbitrarily large radii with arbitrarily little
energy cost; what happened to the UV/IR relation, Equation
\pref{THawk}?  It is probe-dependent.  Objects probing the
flat directions in the moduli space are Dp-branes moving out 
on the Coulomb branch of the effective p+1d superYang-Mills theory;
the UV/IR relation applies to supergravity modes
that do not carry Dp-brane charge, related to single-trace
operators in the gauge theory.%
\footnote{Matrix theory is the extreme example of this
phenomenon -- the light excitations are the center-of-mass motions
of collections of D0-branes, while the single-trace operators
that are perturbations of the supergravity
D0-brane geometry require more energy to excite.}%
\footnote{In the D3-brane gauge theory on $S^3$,
the Coulomb branch is lifted by the scalar field coupling
to the curvature $\RR\phi^2$.  In this case, the
UV/IR relation is more universal.}
The lesson is that
the softest modes of the system are those that
carry Dp-brane charge, and appears to be the key as to when
there is a dual representation of quantum gravity
in terms of a theory without gravity.  The idea
is that when we accumulate a high density of
charged objects, soft excitations occurring inside 
the charge radius must carry the charge, or are built
from composites of charged objects.
Similar phenomena abound in condensed matter systems,
the prime example being the Fermi surface of a conductor,
where all the light excitations are built out of
electronic constituents.
The correspondence fails (\eg\ for D6-branes on $T^6$) when
there are soft excitations that are not built out
of the basic charged constituents; then one
loses control over the gravitational degrees of freedom since
they are not cut off by the regularity of behavior
at high energy in field theory.  This breakdown of the correspondence
is reflected on the gravity side in the negative specific
heat of the corresponding black holes.

%------------ end of article ------------------->>

%% optional
\begin{acknowledgments}
The material in these lectures reflect the thesis work
of my student Vatche Sahakian; I would like to thank him
for an enjoyable and stimulating collaboration.
\end{acknowledgments}

%% appendix optional
%\chapappendix{This is the Appendix Title}
%This is an appendix with a title.

%\chapappendix{}
%This is an appendix without a title.


\begin{chapthebibliography}{99}


%\bibliography{biblio}
%\bibliographystyle{hunsrt}

%\begin{thebibliography}{10}

\bibitem{Aharony:1999ti}
Ofer Aharony, Steven~S. Gubser, Juan Maldacena, Hirosi Ooguri, and Yaron Oz.
\newblock Large N field theories, string theory and gravity.
\newblock 1999, hep-th/9905111.

\bibitem{Horowitz:1997nw}
Gary~T. Horowitz and Joseph Polchinski.
\newblock A correspondence principle for black holes and strings.
\newblock {\em Phys. Rev.}, D55:6189--6197, 1997, hep-th/9612146.

\bibitem{Maldacena:1997re}
Juan Maldacena.
\newblock The large N limit of superconformal field theories and supergravity.
\newblock {\em Adv. Theor. Math. Phys.}, 2:231, 1998, hep-th/9711200.

\bibitem{Susskind:1998dq}
L.~Susskind and E.~Witten.
\newblock The holographic bound in anti-de Sitter space.
\newblock 1998, hep-th/9805114.

\bibitem{Martinec:1998ja}
Emil Martinec and Vatche Sahakian.
\newblock Black holes and the superYang-Mills phase diagram. II.
\newblock {\em Phys. Rev.}, D59:124005, 1999, hep-th/9810224.

\bibitem{Martinec:1999sa}
Emil Martinec and Vatche Sahakian.
\newblock Black holes and five-brane thermodynamics.
\newblock {\em Phys. Rev.}, D60:064002, 1999, hep-th/9901135.

\bibitem{Martinec:1999gw}
Emil Martinec and Vatche Sahakian.
\newblock A note on the thermodynamics of 'little string' theory.
\newblock 1999, hep-th/9906137.

\bibitem{Sahakian:1999gj}
Vatche Sahakian.
\newblock Black holes and thermodynamics of nongravitational theories.
\newblock 1999, hep-th/9906044.

\bibitem{Banks:1997vh}
T.~Banks, W.~Fischler, S.~H. Shenker, and L.~Susskind.
\newblock M theory as a matrix model: A conjecture.
\newblock {\em Phys. Rev.}, D55:5112--5128, 1997, hep-th/9610043.

\bibitem{Banks:1997mn}
T.~Banks.
\newblock Matrix theory.
\newblock {\em Nucl. Phys. Proc. Suppl.}, 67:180, 1998, hep-th/9710231.

\bibitem{Itzhaki:1998dd}
Nissan Itzhaki, Juan~M. Maldacena, Jacob Sonnenschein, and Shimon Yankielowicz.
\newblock Supergravity and the large N limit of theories with sixteen
  supercharges.
\newblock {\em Phys. Rev.}, D58:046004, 1998, hep-th/9802042.

\bibitem{Barbon:1998cr}
J.~L.~F. Barbon, I.~I. Kogan, and E.~Rabinovici.
\newblock On stringy thresholds in SYM/AdS thermodynamics.
\newblock {\em Nucl. Phys.}, B544:104, 1999, hep-th/9809033.

\bibitem{Youm:1997hw}
Donam Youm.
\newblock Black holes and solitons in string theory.
\newblock {\em Phys. Rept.}, 316:1, 1999, hep-th/9710046.

\bibitem{Peet:1998wn}
Amanda~W. Peet and Joseph Polchinski.
\newblock UV/IR relations in AdS dynamics.
\newblock {\em Phys. Rev.}, D59:065011, 1999, hep-th/9809022.

\bibitem{Dijkgraaf:1997hk}
Robbert Dijkgraaf, Erik Verlinde, and Herman Verlinde.
\newblock BPS quantization of the five-brane.
\newblock {\em Nucl. Phys.}, B486:89--113, 1997, hep-th/9604055.

\bibitem{Dijkgraaf:1997nb}
Robbert Dijkgraaf, Erik Verlinde, and Herman Verlinde.
\newblock 5-d black holes and matrix strings.
\newblock {\em Nucl. Phys.}, B506:121, 1997, hep-th/9704018.

\bibitem{Seiberg:1997zk}
Nathan Seiberg.
\newblock New theories in six-dimensions and matrix description of M-theory on
  $T^5$ and $T^5/Z_2$.
\newblock {\em Phys. Lett.}, B408:98--104, 1997, hep-th/9705221.

\bibitem{Giveon:1994fu}
Amit Giveon, Massimo Porrati, and Eliezer Rabinovici.
\newblock Target space duality in string theory.
\newblock {\em Phys. Rept.}, 244:77--202, 1994, hep-th/9401139.

\bibitem{POLCHTASI}
Joseph Polchinski.
\newblock Tasi lectures on D-branes.
\newblock 1996, hep-th/9611050.

\bibitem{Obers:1998fb}
N.~A. Obers and B.~Pioline.
\newblock U-duality and M-theory.
\newblock 1998, hep-th/9809039.

\bibitem{Gregory:1993vy}
R.~Gregory and R.~Laflamme.
\newblock Black strings and $p$-branes are unstable.
\newblock {\em Phys. Rev. Lett.}, 70:2837, 1993, hep-th/9301052.

\bibitem{Horowitz:1998jc}
Gary~T. Horowitz and Joseph Polchinski.
\newblock Selfgravitating fundamental strings.
\newblock {\em Phys. Rev.}, D57:2557--2563, 1998, hep-th/9707170.

\bibitem{Li:1998jy}
Miao Li, Emil Martinec, and Vatche Sahakian.
\newblock Black holes and the SYM phase diagram.
\newblock {\em Phys. Rev.}, D59:044035, 1999, hep-th/9809061.

\bibitem{Banks:1998hz}
T.~Banks, W.~Fischler, I.~R. Klebanov, and L.~Susskind.
\newblock Schwarzschild black holes from matrix theory.
\newblock {\em Phys. Rev. Lett.}, 80:226--229, 1998, hep-th/9709091.

\bibitem{Horowitz:1998fr}
Gary~T. Horowitz and Emil~J. Martinec.
\newblock Comments on black holes in matrix theory.
\newblock {\em Phys. Rev.}, D57:4935--4941, 1998, hep-th/9710217.

\bibitem{Schwarz:1995dk}
John~H. Schwarz.
\newblock An SL(2,Z) multiplet of type IIB superstrings.
\newblock {\em Phys. Lett.}, B360:13--18, 1995, hep-th/9508143.

\bibitem{Aspinwall:1995fw}
Paul~S. Aspinwall.
\newblock Some relationships between dualities in string theory.
\newblock {\em Nucl. Phys. Proc. Suppl.}, 46:30, 1996, hep-th/9508154.

\bibitem{Rozali:1997cb}
Moshe Rozali.
\newblock Matrix theory and U-duality in seven dimensions.
\newblock {\em Phys. Lett.}, B400:260--264, 1997, hep-th/9702136.

\bibitem{Seiberg:1997ad}
Nathan Seiberg.
\newblock Why is the matrix model correct?
\newblock {\em Phys. Rev. Lett.}, 79:3577--3580, 1997, hep-th/9710009.

\bibitem{Motl:1997th}
Lubos Motl.
\newblock Proposals on nonperturbative superstring interactions.
\newblock 1997, hep-th/9701025.

\bibitem{Banks:1997my}
Tom Banks and Nathan Seiberg.
\newblock Strings from matrices.
\newblock {\em Nucl. Phys.}, B497:41--55, 1997, hep-th/9702187.

\bibitem{Dijkgraaf:1997vv}
Robbert Dijkgraaf, Erik Verlinde, and Herman Verlinde.
\newblock Matrix string theory.
\newblock {\em Nucl. Phys.}, B500:43, 1997, hep-th/9703030.

\bibitem{Li:1998ci}
Miao Li and Emil Martinec.
\newblock Probing matrix black holes.
\newblock 1998, hep-th/9801070.

\bibitem{Li:1999qj}
Miao Li.
\newblock Ten-dimensional black hole and the D0-brane threshold bound state.
\newblock {\em Phys. Rev.}, D60:066002, 1999, hep-th/9901158.

\bibitem{Susskind:1997dr}
L.~Susskind.
\newblock Matrix theory black holes and the Gross-Witten transition.
\newblock 1997, hep-th/9805115.

\bibitem{Seiberg:1997ax}
Nathan Seiberg.
\newblock Notes on theories with 16 supercharges.
\newblock {\em Nucl. Phys. Proc. Suppl.}, 67:158, 1998, hep-th/9705117.

\bibitem{Banks:1998dd}
Tom Banks, Michael~R. Douglas, Gary~T. Horowitz, and Emil Martinec.
\newblock AdS dynamics from conformal field theory.
\newblock 1998, hep-th/9808016.

\bibitem{Maldacena:1997tm}
Juan~M. Maldacena.
\newblock Black holes and D-branes.
\newblock {\em Nucl. Phys. Proc. Suppl.}, 61A:111, 1998, hep-th/9705078.

\bibitem{Aharony:1998th}
O.~Aharony, M.~Berkooz, S.~Kachru, N.~Seiberg, and E.~Silverstein.
\newblock Matrix description of interacting theories in six dimensions.
\newblock {\em Adv. Theor. Math. Phys.}, 1:148--157, 1998, hep-th/9707079.

\bibitem{Douglas:1995bn}
Michael~R. Douglas.
\newblock Branes within branes.
\newblock 1995, hep-th/9512077.

\bibitem{Douglas:1996uz}
Michael~R. Douglas.
\newblock Gauge fields and D-branes.
\newblock {\em J. Geom. Phys.}, 28:255, 1998, hep-th/9604198.

\bibitem{Maldacena:1996ds}
Juan~M. Maldacena and Leonard Susskind.
\newblock D-branes and fat black holes.
\newblock {\em Nucl. Phys.}, B475:679--690, 1996, hep-th/9604042.

%\end{thebibliography}

\end{chapthebibliography}
\end{document}